\documentclass[iop]{emulateapj}
\usepackage{appendix}

\begin{document}

\newcommand{\mirlum}{L_{\rm 8}}
\newcommand{\ebmv}{E(B-V)}
\newcommand{\lha}{L(H\alpha)}
\newcommand{\lir}{L_{\rm IR}}
\newcommand{\lbol}{L_{\rm bol}}
\newcommand{\luv}{L_{\rm UV}}
\newcommand{\rs}{{\cal R}}
\newcommand{\ugr}{U_{\rm n}G\rs}
\newcommand{\ks}{K_{\rm s}}
\newcommand{\gmr}{G-\rs}
\newcommand{\hi}{\text{\ion{H}{1}}}
\newcommand{\nhi}{N(\text{\ion{H}{1}})}
\newcommand{\lognhi}{\log[\nhi/{\rm cm}^{-2}]}
\newcommand{\molh}{\text{H}_2}
\newcommand{\nmolh}{N(\molh)}
\newcommand{\lognmolh}{\log[\nmolh/{\rm cm}^{-2}]}

\title{The HDUV Survey: A Revised Assessment of the Relationship
  between UV Slope and Dust Attenuation for High-Redshift Galaxies}

\author{\sc Naveen A. Reddy\altaffilmark{1,10},
Pascal A. Oesch\altaffilmark{2,3},
Rychard J. Bouwens\altaffilmark{4},
Mireia Montes\altaffilmark{3},
Garth D. Illingworth\altaffilmark{5},
Charles C. Steidel\altaffilmark{6},
Pieter G. van Dokkum\altaffilmark{3},
Hakim Atek\altaffilmark{3},
Marcella C. Carollo\altaffilmark{7},
Anna Cibinel\altaffilmark{8},
Brad Holden\altaffilmark{5},
Ivo Labb\'{e}\altaffilmark{4},
Dan Magee\altaffilmark{5},
Laura Morselli\altaffilmark{9},
Erica J. Nelson\altaffilmark{9},
\& Steve Wilkins\altaffilmark{8}
}

\altaffiltext{1}{Department of Physics and Astronomy, University of California, 
Riverside, 900 University Avenue, Riverside, CA 92521, USA; naveenr@ucr.edu}
\altaffiltext{2}{Geneva Observatory, Universit\'{e} de Gen\`{e}ve, Chemin
des Maillettes 51, 1290 Versoix, Switzerland}
\altaffiltext{3}{Yale Center for Astronomy and Astrophysics, Yale University, New
Haven, CT 06511, USA}
\altaffiltext{4}{Leiden Observatory, Leiden University, NL-2300 RA Leiden, Netherlands}
\altaffiltext{5}{UCO/Lick Observatory, University of California, Santa Cruz, 1156
High St, Santa Cruz, CA 95064, USA}
\altaffiltext{6}{Cahill Center for Astronomy and Astrophysics, California Institute of
Technology, MS 249-17, Pasadena, CA 91125, USA}
\altaffiltext{7}{Institute for Astronomy, ETH Zurich, 8092 Zurich, Switzerland}
\altaffiltext{8}{Astronomy Centre, Department of Physics and Astronomy, University 
of Sussex, Brighton, BN1 9QH, UK}
\altaffiltext{9}{Max Planck Institute for Extraterrestrial Physics, Giessenbachstrasse,
85741 Garching bei M\"{u}nchen, Germany}
\altaffiltext{10}{Alfred P. Sloan Research Fellow}

\slugcomment{DRAFT: \today}

\begin{abstract}

We use a newly assembled large sample of 3,545 star-forming galaxies
with secure spectroscopic, grism, and photometric redshifts at
$z=1.5-2.5$ to constrain the relationship between UV slope ($\beta$)
and dust attenuation ($L_{\rm IR}/L_{\rm UV} \equiv {\rm IRX}$).  Our
sample benefits from the combination of deep {\em Hubble} WFC3/UVIS
photometry from the {\em Hubble} Deep UV (HDUV) Legacy survey and
existing photometric data compiled in the 3D-HST survey.  Our sample
significantly extends the range of UV luminosity and $\beta$ probed in
previous samples of UV-selected galaxies, including those as faint as
$M_{1600}=-17.4$ ($\simeq 0.05L^{\ast}_{\rm UV}$) and having $-2.6\la
\beta\la 0.0$.  IRX is measured using stacks of deep {\em
  Herschel}/PACS 100 and $160$\,$\mu$m data, and the results are
compared with predictions of the IRX-$\beta$ relation for different
assumptions of the stellar population model and dust obscuration
curve.  Stellar populations with intrinsically blue UV spectral slopes
necessitate a steeper attenuation curve in order reproduce a given
IRX-$\beta$ relation.  We find that $z=1.5-2.5$ galaxies have an
IRX-$\beta$ relation that is consistent with the predictions for an
SMC extinction curve if we invoke sub-solar ($0.14Z_\odot$)
metallicity models that are currently favored for high-redshift
galaxies, while the commonly assumed starburst attenuation curve
over-predicts the IRX at a given $\beta$ by a factor of $\ga 3$.  The
IRX of high-mass $M_{\ast}> 10^{9.75}$\,$M_\odot$ galaxies is a factor
of $>4$ larger than that of low-mass galaxies, lending support for the
use of stellar mass as a proxy for dust attenuation.  Separate
IRX-$L_{\rm UV}$ relations for galaxies with blue and red $\beta$
conflate to give an average IRX that is roughly constant with UV
luminosity for $L_{\rm UV}\ga 3\times 10^9$\,$L_\odot$.  Thus, the
commonly observed trend of fainter galaxies having bluer $\beta$ may
simply reflect bluer intrinsic UV slopes for such galaxies, rather
than lower dust obscurations.  Taken together with previous studies,
we find that the IRX-$\beta$ distribution for young and low-mass
galaxies at $z\ga 2$ implies a dust curve that is steeper than that of
the SMC, suggesting a lower dust attenuation for these galaxies at a
given $\beta$ relative to older and more massive galaxies.  The lower
dust attenuations and higher ionizing photon output implied by low
metallicity stellar population models point to Lyman continuum
production efficiencies, $\xi_{\rm ion}$, that may be elevated by a
factor of $\approx 2$ relative to the canonical value for $L^{\ast}$
galaxies, aiding in their ability to keep the universe ionized at
$z\sim 2$.

\end{abstract}


\section{INTRODUCTION}
\label{sec:intro}

The ultraviolet (UV) spectral slope, $\beta$, where $f_\lambda\propto
\lambda^\beta$, is by far the most commonly used indicator of dust
obscuration---usually parameterized as the ratio of the infrared-to-UV
luminosity, $L_{\rm IR}/L_{\rm UV}$, or ``IRX'' \citep{calzetti94,
  meurer99}---in moderately reddened high-redshift ($z\ga 1.5$)
star-forming galaxies.  The UV slope can be measured easily from the
same photometry used to select galaxies based on the Lyman break, and
the slope can be used as a proxy for the dust obscuration in galaxies
(e.g., \citealt{calzetti94, meurer99, adelberger00, reddy06a,
  daddi07a, reddy10, overzier11, reddy12a, buat12}) whose dust
emission is otherwise too faint to directly detect in the mid- and
far-infrared (e.g., \citealt{adelberger00, reddy06a}).  Generally,
these studies have indicated that UV-selected star-forming galaxies at
redshifts $1.5\la z\la 3.0$ follow on average the relationship between
UV slope and dust obscuration (i.e., the IRX-$\beta$ relation) found
for local UV starburst galaxies (e.g., \citealt{nandra02, reddy04,
  reddy06a, daddi07a, sklias14}; c.f., \citealt{heinis13, alvarez16}),
though with some deviations that depend on galaxy age \citep{reddy06a,
  siana08, reddy10, buat12}, bolometric luminosity (e.g.,
\citealt{chapman05, reddy06a, casey14b}), stellar mass
\citep{pannella09, reddy10, bouwens16b}, and redshift
\citep{pannella15}.  Unfortunately, typical star-forming ($L^{\ast}$)
galaxies at these redshifts are too faint to directly detect in the
far-infrared.  As such, with the exception of individual lensed galaxy
studies \citep{siana08, siana09, sklias14, watson15, dessauges16},
most investigations that have explored the relation between UV slope
and dust obscuration for moderately reddened galaxies have relied on
stacking relatively small numbers of objects and/or used shorter
wavelength emission---such as that arising from polycyclic aromatic
hydrocarbons (PAHs)---to infer infrared luminosities.

New avenues of exploring the dustiness of high-redshift galaxies have
been made possible with facilities such as the Atacama Large
Millimeter Array (ALMA), allowing for direct measurements of either
the dust continuum or far-IR spectral features for more typical
star-forming galaxies in the distant universe \citep{carilli13, dunlop17}.
Additionally, the advent of large-scale rest-optical spectroscopic
surveys of intermediate-redshift galaxies at $1.4\la z\la 2.6$---such
as the 3D-HST \citep{vandokkum13}, the MOSFIRE Deep Evolution Field
(MOSDEF; \citealt{kriek15}), and the Keck Baryonic Structure surveys
(KBSS; \citealt{steidel14})---have enabled measurements of obscuration
in individual high-redshift star-forming galaxies using Balmer
recombination lines (e.g., \citealt{price14, reddy15, nelson16}).  While these
nebular line measurements will be possible in the near future for
$z\ga 3$ galaxies with the {\em James Webb Space Telescope} ({\em
  JWST}), the limited lifetime of this facility and the targeted
nature of both ALMA far-IR and {\em JWST} near- and mid-IR
observations means that the UV slope will remain the only easily
accessible proxy for dust obscuration for large numbers of {\em
  individual} typical galaxies at $z\ga 3$ in the foreseeable future.

Despite the widespread use of the UV slope to infer dust attenuation,
there are several complications associated with its use. First, the UV
slope is sensitive to metallicity and star-formation history (e.g.,
\citealt{kong04, seibert05, johnson07b, dale09, munoz09, reddy10, wilkins11,
  boquien12, reddy12b, schaerer13, wilkins13, grasha13, zeimann15}).  Second,
there is evidence that the relationship between UV slope and dust
obscuration depends on stellar mass and/or age (e.g.,
\citealt{reddy06a, buat12, zeimann15, bouwens16b}), perhaps reflecting
variations in the shape of the attenuation curve.  Third, the
measurement of the UV slope may be complicated by the presence of the
2175\,\AA\, absorption feature \citep{noll09, buat11, kriek13, buat12,
  reddy15}.  Fourth, as noted above, independent inferences of the
dust attenuation in faint galaxies typically involve stacking mid- and
far-IR data, but such stacking masks the scatter in the relationship
between UV slope and obscuration.  Quantifying this scatter can
elucidate the degree to which the attenuation curve may vary from
galaxy-to-galaxy, or highlight the sensitivity of the UV slope to
factors other than dust obscuration.  In general, the effects of age,
metallicity, and star-formation history on the UV slope may become
important for ultra-faint galaxies at high redshift which have been
suggested to undergo bursty star formation (e.g., \citealt{weisz12,
  hopkins14, dominguez15, guo16, sparre17, faucher17}).

Obtaining direct constraints on the dust obscuration of UV-faint
galaxies is an important step in evaluating the viability of the UV
slope to trace dustiness, quantifying the bolometric luminosities of
ultra-faint galaxies and their contribution to the global SFR and
stellar mass densities, assessing possible variations in the dust
obscuration curve over a larger dynamic range of galaxy
characteristics (e.g., star-formation rate, stellar mass, age,
metallicity, etc.), and discerning the degree to which the UV slope
may be affected by short timescale variations in star-formation rate.

Separately, recent advances in stellar populations models that include
realistic treatments of stellar mass loss, rotation, and multiplicity
\citep{eldridge09, brott11, levesque12, leitherer14} can result in
additional dust heating from ionizing and/or recombination photons.
Moreover, the intrinsic UV spectral slopes of high-redshift galaxies
with lower stellar metallicities may be substantially bluer
\citep{schaerer13, sklias14, alavi14, cullen17} than what has been
typically assumed in studies of the IRX-$\beta$ relation.  Thus, it
seems timely to re-evaluate the IRX-$\beta$ relation in light of these
issues.

With this in mind, we use a newly assembled large sample of galaxies
with secure spectroscopic or photometric redshifts at $1.5\le z\le
2.5$ in the GOODS-North and GOODS-South fields 
to investigate the correlation between UV slope and dust obscuration.
Our sample takes advantage of newly acquired {\em Hubble} UVIS F275W
and F336W imaging from the HDUV survey (Oesch et~al. 2017, submitted)
which aids in determining photometric redshifts when combined with
existing 3D-HST photometric data.  This large sample enables precise
measurements of dust obscuration through the stacking of far-infrared
images from the {\em Herschel Space Observatory}, and also enables
stacking in multiple bins of other galaxy properties (e.g., stellar
mass, UV luminosity) to investigate the scatter in the IRX-$\beta$
relation.
We also consider the newest stellar population models---those which
may be more appropriate in describing very high-redshift ($z\ga 2$)
galaxies---in interpreting the relationship between UV slope and
obscuration.  

\begin{deluxetable*}{lc}
\tabletypesize{\footnotesize}
\tablewidth{0pc}
\tablecaption{Sample Characteristics}
\tablehead{
\colhead{Property} &
\colhead{Value}}
\startdata
Fields & GOODS-N, GOODS-S \\
Total area & $\sim 329$\,arcmin$^2$ \\
Area with HDUV imaging & $\sim 100$\,arcmin$^{2}$ \\
UV/Optical photometry & 3D-HST Catalogs\tablenotemark{a} and HDUV\tablenotemark{b} F275W and F336W \\
Mid-IR imaging & {\em Spitzer} GOODS Imaging Program\tablenotemark{c} \\
Far-IR imaging & GOODS-{\em Herschel}\tablenotemark{d} and PEP\tablenotemark{e} Surveys \\
Optical depth of sample & $H \simeq 27$ \\
UV depth of sample & $m_{\rm UV} \simeq 27$ \\
Total number of galaxies & 4,078 \\
Number of galaxies with far-IR coverage & 3,569 \\
Final number (excl. far-IR-detected objects) & 3,545 \\
$\beta$ Range & $-2.55\le \beta \le 1.05$ ($\langle\beta\rangle = -1.71$)
\enddata
\tablenotetext{a}{\citet{skelton14}.}
\tablenotetext{b}{Oesch et~al., submitted.}
\tablenotetext{c}{PI: Dickinson.}
\tablenotetext{d}{\citet{elbaz11}.}
\tablenotetext{e}{PI: Lutz, \citet{magnelli13}.}
\label{tab:sample}
\end{deluxetable*}

The outline of this paper is as follows.  In Section~\ref{sec:sample},
we discuss the selection and modeling of stellar populations of
galaxies used in this study.  The methodology used for stacking the
mid- and far-IR {\em Spitzer} and {\em Herschel} data is discussed in
Section~\ref{sec:stacking}.  In Section~\ref{sec:predirx}, we
calculate the predicted relationships between IRX and $\beta$ for
different attenuation/extinction curves using energy balance
arguments.  These predictions are compared to our (as well as
literature) stacked measurements of IRX in
Section~\ref{sec:discussion}.  In this section, we also consider the
variation of IRX with stellar masses, UV luminosities, and the ages of
galaxies, as well as the implications of our results for modeling the
stellar populations and inferring the ionizing efficiencies of
high-redshift galaxies.  AB magnitudes are assumed throughout
\citep{oke83}, and we use a \citet{chabrier03} initial mass function
(IMF) unless stated otherwise. We adopt a cosmology with
$H_{0}=70$\,km\,s$^{-1}$\,Mpc$^{-1}$, $\Omega_{\Lambda}=0.7$, and
$\Omega_{\rm m}=0.3$.

\section{SAMPLE AND IR IMAGING}
\label{sec:sample}

\subsection{Parent Sample}

A few basic properties of our sample are summarized in
Table~\ref{tab:sample}.  Our sample of galaxies was constructed by
combined the publicly-available ground- and space-based photometry
compiled by the 3D-HST survey \citep{skelton14} with newly obtained
imaging from the {\em Hubble} Deep UV (HDUV) Legacy Survey (GO-13871;
Oesch et~al. 2017, submitted).  The HDUV survey imaged the two GOODS
fields in the F275W and F336W bands to depths of $\simeq 27.5$ and
$27.9$\,mag, respectively ($5\sigma$; $0\farcs4$ diameter aperture),
with the UVIS channel of the {\em Hubble Space Telescope} WFC3
instrument.  A significant benefit of the HDUV imaging is that it
allows for the Lyman break selection of galaxies to fainter UV
luminosities and lower redshifts than possible from ground-based
surveys (Oesch et~al. 2017, submitted), and builds upon previous
efforts to use deep UVIS imaging to select Lyman break galaxies at
$z\sim 2$ \citep{hathi10, windhorst11}.  The reduced UVIS images,
covering $\approx 100$\,arcmin$^{2}$, include previous imaging
obtained by the CANDELS \citep{koekemoer11} and UVUDF surveys
\citep{teplitz13, rafelski15}.

\subsection{Photometry and Stellar Population Parameters}

Source Extractor \citep{bertin96} was used to measure photometry on
the UVIS images using the detection maps for the combined
F125W$+$F140W$+$F160W images, as was done for the 3D-HST photometric
catalogs \citep{skelton14}.  The publicly-available 3D-HST photometric
catalogs were then updated with the HDUV photometry---i.e., such that
the updated catalogs contain updated photometry for objects lying in
the HDUV pointings as well as the original set photometry for objects
lying outside the HDUV pointings.  This combined dataset was then used
to calculate photometric redshifts using EAZY \citep{brammer08} and
determine stellar population parameters (e.g., stellar mass) using
FAST \citep{kriek09}.  Where available, grism and external
spectroscopic redshifts were used in lieu of the photometric redshifts
when fitting for the stellar populations.  These external
spectroscopic redshifts are provided in the 3D-HST catalogs
\citep{momcheva16}.  We also included 759 spectroscopic redshifts for
galaxies observed during the 2012B-2015A semesters of the MOSDEF
survey \citep{kriek15}.  

For the stellar population modeling, we adopted the \citet{conroy10}
stellar population models for $Z=0.019$\,Z$_\odot$, a delayed-$\tau$
star-formation history with $8.0\le \log[\tau/{\rm yr}] \le 10.0$, a
\citet{chabrier03} initial mass function (IMF), and the
\citet{calzetti00} dust attenuation curve with $0.0\le A_{\rm V}\le
4.0$.\footnote{Below, we consider the effect of stellar population age
  on the IRX-$\beta$ relations.  In that context, the ages derived for
  the vast majority of galaxies in our sample are within
  $\delta\log[{\rm Age/yr}] \simeq 0.1$\,dex to those derived assuming
  a constant star-formation history.}  We imposed a minimum age of
$40$\,Myr based on the typical dynamical timescale for $z\sim 2$
galaxies \citep{reddy12b}.

The UV slope for each galaxy was calculated both by (a) fitting a
power law through the broadband photometry, including only bands lying
redward of the Lyman break and blueward of rest-frame $2600$\,$\mu$m;
and (b) fitting a power law through the best-fit SED points that lie
in wavelength windows spanning rest-frame $1268\le \lambda\le
2580$\,\AA, as defined in \citet{calzetti94}.  Method (a) includes a
more conservative estimate for the errors in $\beta$, but generally
the two methods yielded values of the UV slope for a given galaxy that
were within $\delta\beta \simeq 0.1$ of each other.  We adopted the
$\beta$ calculated using method (a) for the remainder of our analysis,
and note that in Section~\ref{sec:predirx}, we consider the value of $\beta$
using windows lying strictly blueward of $\approx 1800$\,\AA.

\begin{figure*}
\epsscale{1.15}
\plotone{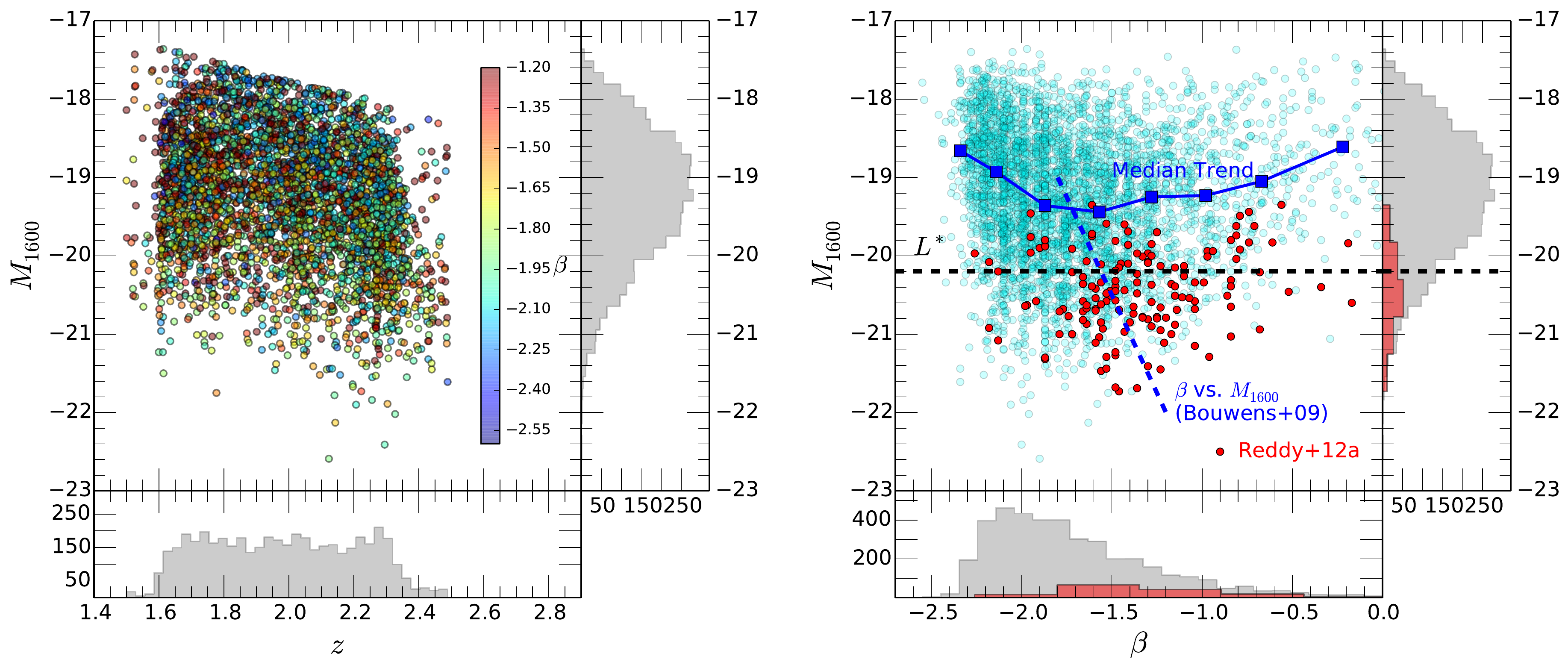}
\caption{Redshift ($z$), absolute magnitude ($M_{1600}$), and UV slope
  ($\beta$) distributions of the 3,545 galaxies in our sample,
  color-coded by $\beta$ (left panel) and denoted by cyan symbols
  (right panel).  For comparison, the distributions for the 124
  UV-selected galaxies from \citet{reddy12a} are indicated by the red
  symbols and histograms in the right panel, while the black dashed
  line denotes the value of $M^{\ast}_{1500} \simeq -20.2$ at $z\sim
  1.9$ from \citet{oesch10}.  The present HDUV+3D-HST sample is a
  factor of $\approx 30\times$ larger than that of \citet{reddy12a},
  and includes galaxies over broader ranges of $M_{1600}$ and $\beta$,
  particularly at fainter magnitudes and bluer UV slopes.  The blue
  squares and solid line in the right panel indicate the median
  relationship between $\beta$ and $M_{1600}$ for our sample, compared
  with the mean relationship (blue dashed line) found for 168
  $U$-dropouts at $z\sim 2.5$ (note this is the upper bound in
  redshift for our sample) from \citet{bouwens09}.  Removing the
  $m_{(1+z)\times 1700}\le 27.0$ requirement imposed to construct our
  sample (Section~\ref{sec:sample}) would allow for a larger number of
  faint galaxies with redder UV slopes to be selected, increasing the
  number density of objects in the upper right-hand region of the
  right panel.}
\label{fig:zmag}
\end{figure*}

\subsection{Criteria for Final Sample}

The photometric catalogs, along with those containing the redshifts
and stellar population parameters, were used to select galaxies based
on the following criteria.  First, the object must have a Source
Extractor ``class star'' parameter $< 0.95$, or observed-frame $U-J$
and $J-K$ colors that reside in the region occupied by star-forming
galaxies as defined by \citep{skelton14}---these criteria ensure the
removal of stars.  

Second, the galaxy must have a spectroscopic or grism redshift, or
$95\%$ confidence intervals in the photometric redshift, that lie in
the range $1.5\le z\le 2.5$.  Note that the high photometric redshift
confidence intervals required for inclusion in our sample naturally
selects those objects with $H\la 27$.

Third, the object must not have a match in X-ray AGN catalogs compiled
for the GOODS-North and GOODS-South fields (e.g., \citealt{shao10,
  xue11}).  Additionally, we use the \citet{donley12} {\em Spitzer}
IRAC selection to isolate any infrared-bright AGN.  While the X-ray
and IRAC selections may not identify {\em all} AGN at the redshifts of
interest, they are likely to isolate those AGN that may significantly
influence our stacked far-IR measurements.  

Fourth, the object must not have rest-frame $U-V$ and $V-J$ colors
that classify it as a quiescent galaxy \citep{williams09, skelton14}.
The object is further required to have a specific star-formation rate
sSFR$\ga 0.1$\,Gyr$^{-1}$.  These criteria safeguard against the
inclusion of galaxies where $\beta$ may be red due to the contribution
of older stars to the near-UV continuum, or where dust heating by
older stars may become significant.  

Fifth, to ensure that the sample is not biased towards objects with
red $U-H$ colors at faint $U$ magnitudes (owing to the limit in
$H$-band magnitude mentioned previously), the galaxy must have an
apparent magnitude at $[1+z]\times 1600$\,\AA\, of $\le 27.0$\,mag.
This limit still allows us to include galaxies with absolute
magnitudes as faint as $M_{\rm 1600}\simeq -17.4$.  These criteria
result in a sample of 4,078 galaxies.

\subsection{Spitzer and Herschel Imaging}

We used the publicly available {\em Spitzer}/MIPS 24\,$\mu$m and {\em
  Herschel}/PACS 100 and 160\,$\mu$m data in the two GOODS fields for
our analysis.  The 24\,$\mu$m data come from the {\em Spitzer} GOODS
imaging program (PI: Dickinson), and trace the dust-sensitive
rest-frame $7.7$\,$\mu$m emission feature for galaxies at $1.5\le z\le
2.5$ (e.g., \citealt{reddy06a}).  The observed $24$\,$\mu$m fluxes of
$z\sim 2$ galaxies have been used extensively in the past to derive
infrared luminosities ($L_{\rm IR}$) given the superior sensitivity of
these data to dust emission when compared with observations taken at
longer wavelengths (roughly a factor of three times more sensitive
than {\em Herschel}/PACS to galaxies of a given $L_{\rm IR}$ at $z\sim
2$; \citealt{elbaz11}).  However, a number of observations have
highlighted the strong variation in $L_{\rm 7.7}/L_{\rm IR}$ with
star-formation rate \citep{rieke09, shipley16}, star-formation-rate
surface density \citep{elbaz11}, and gas-phase metallicity and
ionization parameter at high-redshift \citep{shivaei16}.  As such,
while we stacked the $24$\,$\mu$m data for galaxies in our sample, we
did not consider these measurements when calculating $L_{\rm IR}$.  In
Appendix~\ref{sec:l8lir}, we consider further the variation in $L_{\rm
  7.7}/L_{\rm IR}$ with other galaxy characteristics.

The {\em Herschel} data come from the GOODS-{\em Herschel} Open Time
Key Program \citep{elbaz11} and the PACS Evolutionary Probe (PEP)
Survey (PI: Lutz; \citealt{magnelli13}), and probe the rest-frame
$\simeq 30-65$\,$\mu$m dust continuum emission for galaxies at $1.5\le
z\le 2.5$.  We chose not to use the SPIRE data given the much coarser
spatial resolution of these data (FWHM$\ga 18\arcsec$) relative to the
$24$\,$\mu$m (FWHM$\simeq 5\farcs 4$), $100$\,$\mu$m (FWHM$\simeq
6\farcs 7$), and $160$\,$\mu$m (FWHM$\simeq 11\arcsec$) data.  The
pixel scales of the 24, 100, and 160\,$\mu$m images are $1\farcs 2$,
$1\farcs 2$, and $2\farcs 4$, respectively.  As noted above, only the
$100$ and $160$\,$\mu$m data are used to calculate $L_{\rm IR}$.

Of the 4,078 galaxies in the sample discussed above, 3,569 lie within
the portions of the {\em Herschel} imaging that are $80\%$ complete to
flux levels of 1.7 and 5.5\,mJy for the 100 and 160\,$\mu$m maps in
GOODS-N, respectively, and 1.3 and 3.9\,mJy for the 100 and
160\,$\mu$m maps in GOODS-S, respectively.  Of these galaxies, 24 (or
$0.67\%$) are directly detected with signal-to-noise $S/N>3$ in either
the $100$ or $160$\,$\mu$m images.  As we are primarily concerned with
constraining the IRX-$\beta$ relation for {\em moderately} reddened
galaxies, we removed all directly-detected {\em Herschel} objects from
our sample---the latter are very dusty star-forming galaxies at the
redshifts of interest with $L_{\rm IR}\ga 10^{12}$\,$L_\odot$.  The
very low frequency of infrared-luminous objects among UV-faint
galaxies in general could have been anticipated from the implied low
number density of $L_{\rm IR}\ga 10^{12}$\,$L_\odot$ objects from the
IR luminosity function \citep{reddy08, magnelli13} and the high number
density of UV-faint galaxies inferred from the UV luminosity function
\citep{reddy08, reddy09, alavi16} at $z\sim2$.  The inclusion of such
dusty galaxies does not significantly affect our stacking analysis
owing to the very small number of such objects.  Excluding these dusty
galaxies, our final sample consists of 3,545 galaxies with the
redshift and absolute magnitude distributions shown in
Figure~\ref{fig:zmag}.

\subsection{Summary of Sample}

To summarize, we have combined HDUV UVIS and 3D-HST catalogued
photometry to constrain photometric redshifts for galaxies in the
GOODS fields and isolate those star-forming galaxies with redshifts
$z=1.5-2.5$ down to a limiting near-IR magnitude of $\simeq 27$\,AB
(Table~\ref{tab:sample}).  All galaxies are significantly detected
(with $S/N>3$) down to an observed optical (rest-frame UV) magnitude
of $27$\,AB.  Our sample includes objects with spectroscopic redshifts
in the aforementioned range wherever possible.  This sample is then
used as a basis for stacking deep {\em Herschel} data, as discussed in
the next section.

One of the most beneficial attributes of our sample is that it
contains the largest number of UV-faint galaxies---extending up to
$\approx 3$ magnitudes fainter than the characteristic absolute
magnitude at $z\sim 2.3$ ($M^{\ast}_{1700}=-20.70$; \citealt{reddy09})
and $z\sim 1.9$ ($M^{\ast}_{1500}=-20.16$; \citealt{oesch10})---with
robust redshifts at $1.5\le z\le 2.5$ assembled to date
(Figure~\ref{fig:zmag}).  The general faintness of galaxies in our
sample is underscored by their very low detection rate ($S/N>3$) at
$24$\,$\mu$m---85 of 3,545 galaxies, or $\approx 2.4\%$---compared to
the $\approx 40\%$ detection rate for rest-frame UV-selected galaxies
with ${\cal R}\le 25.5$ \citep{reddy10}.  Consequently, unlike most
previous efforts using ground-based UV-selected samples of limited
depth, the present sample presents a unique opportunity to evaluate
the IRX-$\beta$ relation for the analogs of the very faint galaxies
that dominate the UV and bolometric luminosity densities at $z\gg3$
(e.g., \citealt{reddy08, smit12}), but for which direct constraints on
their infrared luminosities are difficult to obtain.

\begin{deluxetable*}{lccccccccc}
\tabletypesize{\footnotesize}
\tablewidth{0pc}
\tablecaption{Stacked Fluxes and Infrared and UV Luminosities}
\tablehead{
\colhead{} &
\colhead{} &
\colhead{} &
\colhead{} &
\colhead{$\langle f_{24}\rangle$\tablenotemark{c}} &
\colhead{$\langle f_{100}\rangle$\tablenotemark{c}} &
\colhead{$\langle f_{160}\rangle$\tablenotemark{c}} &
\colhead{$\langle L_{7.7}\rangle$\tablenotemark{d}} &
\colhead{$\langle L_{\rm IR}\rangle$\tablenotemark{d}} &
\colhead{$\langle L_{\rm UV}\rangle$\tablenotemark{d}} \\
\colhead{Sample} &
\colhead{$N$\tablenotemark{a}} &
\colhead{$\langle z\rangle$\tablenotemark{b}} &
\colhead{$\langle \beta\rangle$\tablenotemark{b}} &
\colhead{[$\mu$Jy]} &
\colhead{[$\mu$Jy]} &
\colhead{[$\mu$Jy]} &
\colhead{[$10^{10}$\,$L_\odot$]} &
\colhead{[$10^{10}$\,$L_\odot$]} &
\colhead{[$10^{10}$\,$L_\odot$]}}
\startdata
{\bf All} & 3545 & 1.94 & -1.71 & $1.54\pm0.14$ & $29\pm6$ & $62\pm17$ & $0.26\pm0.03$ & $2.1\pm0.4$ & 0.80 \\
\\
{\bf $M_{1600}$ bins:} \\
$M_{1600} \le -21$ & 81 & 2.12 & -1.74 & $4.83\pm0.96$ & $177\pm30$ & $377\pm93$ & $1.00\pm0.20$ & $17.1\pm2.4$ & 6.73 \\
$-21<M_{1600}\le -20$ & 575 & 2.07 & -1.68 & $4.37\pm0.28$ & $87\pm13$ & $171\pm43$ & $0.86\pm0.06$ & $7.6\pm1.0$ & 2.92 \\
$-20<M_{1600}\le -19$ & 1390 & 1.99 & -1.67 & $2.33\pm0.20$ & $38\pm8$ & $84\pm25$ & $0.41\pm0.03$ & $3.1\pm0.6$ & 1.26 \\
$M_{1600}>-19$ & 1499 & 1.92 & -1.72 & $1.00\pm0.16$ & $31\pm9$ & $54\pm24$ & $0.16\pm0.03$ & $2.0\pm0.5$ & 0.48 \\
\\
{\bf $\beta$ bins:} \\
$\beta \le -1.70$ & 2084 & 1.96 & -2.04 & $0.52\pm0.16$ & $5\pm7$ & $21\pm18$ & $0.09\pm0.03$ & $<1.4$ & 0.77 \\
$-1.70<\beta\le -1.40$ & 722 & 1.92 & -1.56 & $1.89\pm0.41$ & $43\pm13$ & $86\pm37$ & $0.31\pm0.07$ & $2.9\pm0.7$ & 0.95 \\
$-1.40<\beta\le -1.10$ & 345 & 1.94 & -1.26 & $3.92\pm0.55$ & $52\pm18$ & $103\pm56$ & $0.65\pm0.09$ & $3.7\pm1.1$ & 0.93 \\
$-1.10<\beta\le -0.80$ & 205 & 1.93 & -0.97 & $7.07\pm0.53$ & $80\pm25$ & $173\pm73$ & $1.15\pm0.09$ & $5.7\pm1.4$ & 0.81 \\
$\beta>-0.80$ & 189 & 1.90 & -0.31 & $5.09\pm0.62$ & $167\pm23$ & $340\pm63$ & $0.80\pm0.10$ & $11.0\pm1.2$ & 0.59 \\
\\
{\bf $M_{1600}$ \& $\beta$ bins:}\\
$M_{1600}\le -19$ $+$ $\beta \le -1.4$ & 1616 & 2.01 & -1.86 & $1.86\pm0.21$ & $25\pm9$ & $51\pm21$ & $0.33\pm0.04$ & $1.9\pm0.5$ & 1.58 \\
$M_{1600}\le -19$ $+$ $\beta > -1.4$ & 430 & 1.97 & -1.02 & $7.20\pm0.50$ & $117\pm19$ & $288\pm46$ & $1.25\pm0.09$ & $9.5\pm1.0$ & 1.47 \\
$M_{1600}> -19$ $+$ $\beta \le -1.4$ & 1190 & 1.92 & -1.97 & $0.36\pm0.21$ & $13\pm7$ & $26\pm23$ & $0.06\pm0.03$ & $<1.0$ & 0.48 \\
$M_{1600}> -19$ $+$ $\beta > -1.4$ & 309 & 1.90 & -0.79 & $3.11\pm0.38$ & $95\pm19$ & $176\pm75$ & $0.49\pm0.06$ & $6.3\pm1.2$ & 0.48 \\
\\
{\bf Stellar Mass \& $\beta$ bins:} \\
$\log[M_{\ast}/{\rm M}_\odot]\le 9.75$ & 2571 & 1.94 & -1.88 & $0.75\pm0.14$ & $10\pm7$ & $17\pm20$ & $0.13\pm0.03$ & $<1.2$ & 0.71 \\
$\,\,\,\,\,\,+\beta\le -1.4$ & 2385 & 1.94 & -1.95 & $0.63\pm0.19$ & $11\pm6$ & $28\pm15$ & $0.10\pm0.03$ & $<1.0$ & 0.72 \\
$\,\,\,\,\,\,+\beta>-1.4$ & 186 & 1.89 & -1.12 & $2.95\pm0.76$ & $19\pm25$ & $72\pm73$ & $0.47\pm0.12$ & $<4.0$ & 0.57 \\
$\log[M_{\ast}/{\rm M}_\odot]>9.75$ & 974 & 1.96 & -0.92 & $4.93\pm0.40$ & $111\pm 12$ & $229\pm36$ & $0.84\pm0.07$ & $8.3\pm0.7$ & 1.22 \\
$\,\,\,\,\,\,+\beta\le -1.4$ & 421 & 2.04 & -1.61 & $4.22\pm0.42$ & $59\pm14$ & $118\pm46$ & $0.80\pm0.07$ & $5.1\pm1.1$ & 2.26 \\
$\,\,\,\,\,\,+\beta>-1.4$ & 553 & 1.94 & -0.72 & $5.23\pm0.48$ & $132\pm14$ & $263\pm44$ & $0.87\pm0.09$ & $9.4\pm0.8$ & 0.90 \\
\\
{\bf Age bins:} \\
$\log[{\rm Age}/{\rm yr}]\le 8.00$ & 81 & 1.96 & -1.49 & $0.32\pm0.92$ & $62\pm39$ & $135\pm91$ & $<0.51$ & $<6.3$ & 0.55 \\
$\log[{\rm Age}/{\rm yr}]>8.00$ & 3464 & 1.94 & -1.71 & $1.43\pm0.22$ & $25\pm6$ & $52\pm19$ & $0.23\pm0.04$ & $1.8\pm0.4$ & 0.81 
\enddata
\tablenotetext{a}{Number of objects in the stack.}
\tablenotetext{b}{Mean redshift and UV slope of objects in the stack.}
\tablenotetext{c}{Stacked 24, 100, and 160\,$\mu$m fluxes.}
\tablenotetext{d}{Stacked 8\,$\mu$m and total infrared luminosities, and the mean UV luminosity
of objects in the stack.}
\label{tab:stackedresults}
\end{deluxetable*}

\section{STACKING METHODOLOGY}
\label{sec:stacking}

To mitigate any systematics in the stacked fluxes due to bright
objects proximate to the galaxies in our sample, we performed the
stacking on residual images that were constructed as
follows.\footnote{As discussed in \citet{reddy12a}, stacking on the
  science images themselves yields results similar to those obtained
  by stacking on the residual images.}  We used the $24$\,$\mu$m
catalogs and point spread functions (PSFs) included in the GOODS-{\em
  Herschel} data release to subtract all objects with $S/N>3$ in the
$24$\,$\mu$m images, with the exception of the 85 objects in our
sample that are directly detected at $24$\,$\mu$m.  Objects with
$S/N>3$ in the 24\,$\mu$m images were used as priors to fit and
subtract objects with $S/N>3$ in the 100 and 160\,$\mu$m images.  The
result is a set of residual images at 24, 100, and 160\,$\mu$m for
both GOODS fields.

For each galaxy contributing to the stack, we extracted from the 24,
100, and 160\,$\mu$m residual images regions of $41\times 41$,
$52\times 52$, and $52\times 52$ pixels, respectively, centered on the
galaxy.  The sub-images were then divided by the UV luminosity,
$L_{\rm UV}=\nu L_\nu$ at $1600$\,\AA, of the galaxy, and these
normalized sub-images for each band were then averaged together using
3\,$\sigma$ clipping for all the objects in the stack.  We performed
PSF photometry on the stacked images to measure the fluxes.  Because
the images are normalized by $L_{\rm UV}$, the stacked fluxes are
directly proportional to the average IRX.  The corresponding weighted
average fluxes in each band ($\langle f_{24}\rangle$, $\langle
f_{100}\rangle$, and $\langle f_{160}\rangle$), where the weights are
$1/L_{\rm UV}$, were computed by multiplying the stacked fluxes by the
weighted average UV luminosity of galaxies in the stack.  The
measurement uncertainties of these fluxes were calculated as the
1\,$\sigma$ dispersion in the fluxes obtained by fitting PSFs at
random positions in the stacked images, avoiding the stacked signal
itself.

While stacking on residual images aids in minimizing the contribution
of bright nearby objects to the stacked fluxes, this method will not
account for objects that are blended with the galaxies of interest in
the {\em Herschel}/PACS imaging.  This presents a particular challenge
in our case, where the galaxies are selected from {\em HST}
photometry, as a single galaxy (e.g., as observed from the ground) may
be resolved with {\em HST} into several subcomponents, each of which
is of course unresolved in the {\em Herschel} imaging but each of
which will contribute to the stacked flux.  Galaxies that are resolved
into multiple subcomponents will contribute more than once to the
stack, resulting in an over-estimate of the stacked far-IR flux.  This
effect is compounded by that of separate galaxies contributing more
than once to the stack if they happen to be blended at the {\em
  Herschel}/PACS resolution.  This bias was quantified as follows.

For a given band, we used the PSF to generate $N$ galaxies, where $N$
is the number of galaxies in the stack, each having a flux equal to
the weighted average flux of the stacked signal.  These simulated
galaxies were added to the residual image at locations that were
shifted from those of the real galaxies by offsets $\delta x$ and
$\delta y$ in the x- and y-directions, respectively, where the offsets
were chosen randomly.  This ensures that the spatial distribution of
the simulated galaxies is identical to that of the real galaxies.  We
then stacked at the locations of the simulated galaxies and compared
the simulated and recovered stacked fluxes.  This was done 100 times,
each time with a different pair of (randomly chosen) $\delta x$ and
$\delta y$.  The average ratio of the simulated and recovered fluxes,
or the bias factor, from these 100 simulations varied from $\approx
0.60-0.90$, depending on the number of galaxies contributing to the
stack and the particular band.  These simulations were performed for
every band and for every stack in our analysis, and the stacked fluxes
of the galaxies in our sample were multiplied by the bias factors
calculated from these simulations.

To further investigate this bias, we also stacked all galaxies in our
sample that had no {\em HST}-detected object within $3\farcs 35$,
corresponding to the half-width at half-maximum of the {\em
  Herschel}/PACS 100\,$\mu$m PSF.  While this criterion severely
restricts the size of the sample to only 465 objects, it allowed us to
verify the bias factors derived from our simulations.  Stacking these
465 objects yielded weighted average fluxes at $24$, $100$, and
$160$\,$\mu$m that are within 1\,$\sigma$ of the those values obtained
for the entire sample once the bias factors are
applied.\footnote{While the 160\,$\mu$m PSF has a half-width at
  half-maximum that is larger than the exclusion radius of $3\farcs
  35$, the agreement in the average $f_{100}/f_{160}$ ratio, or
  far-infrared color, between the stack of the full sample and that of
  the 465 galaxies suggests that the bias factors also recover
  successfully the average 160\,$\mu$m stacked flux.}

Infrared luminosities were calculated by fitting the \citet{elbaz11}
``main sequence'' dust template to the stacked $\langle
f_{100}\rangle$ and $\langle f_{160}\rangle$ fluxes.  We chose this
particular template as it provided the best match to the observed
infrared colors $f_{100}/f_{160}$ of the stacks, though we note that
the adoption of other templates (e.g., \citealt{chary01, dale02,
  rieke09}) results in $L_{\rm IR}$ that vary by no more than $\approx
50\%$ from the ones calculated here (see \citealt{reddy12a} for a
detailed comparison of $L_{\rm IR}$ computed using different dust
templates).  Upper limits in $L_{\rm IR}$ are quoted in cases where
$L_{\rm IR}$ divided by the modeled uncertainty is $>3$.  In a few
instances, $\sim 2\,\sigma$ detections of both the $100$ and
$160$\,$\mu$m stacked fluxes yield a modeled $L_{\rm IR}$ that is
significant at the $3\sigma$ level.

The mean UV slope of objects contributing to the stack was computed as
a weighted average of the UV slopes of individual objects where,
again, the weights are $1/L_{\rm UV}$.  These same weights were also
applied when calculating the weighted average redshift, absolute
magnitude, stellar mass, and age of objects contributing to the stack.
Table~\ref{tab:stackedresults} lists the average galaxy properties and
fluxes for each stack performed in our study.

\section{PREDICTED IRX-$\beta$ RELATIONS}
\label{sec:predirx}

We calculated the relationship between IRX and $\beta$ using the
recently developed ``Binary Population and Spectral Synthesis''
(BPASS) models \citep{eldridge12, stanway16} with a stellar
metallicity of $Z=0.14Z_\odot$ on the current abundance scale
\citep{asplund09} and a two power-law IMF with $\alpha = 2.35$ for
$M_{\ast} > 0.5$\,$M_\odot$ and $\alpha= 1.30$ for $0.1\le M_{\ast}\le
0.5$\,$M_\odot$.  We assumed a constant star formation with an age of
$100$\,Myr and included nebular continuum emission.  This particular
BPASS model (what we refer to as our ``fiducial'' model) is found to
best reproduce simultaneously the rest-frame far-UV continuum,
stellar, and nebular lines, and the rest-frame optical nebular
emission line strengths measured for galaxies at $z\sim 2$
\citep{steidel16}.  Two salient features of this model are the very
blue intrinsic UV continuum slope $\beta_0 \simeq -2.62$ relative to
that assumed in the \citet{meurer99} calibration of the IRX-$\beta$
relation ($\beta_0 = -2.23$), and the larger number of ionizing
photons per unit star-formation-rate (i.e., $\approx 20\%$ larger than
those of single star models with no stellar rotation;
\citealt{stanway16}) that are potentially available for heating dust.
For comparison, the BPASS model for the same metallicity with a
constant star-formation history and an age of $300$\,Myr (the median
for the sample considered here, and similar to the mean age of $z\sim
2$ UV-selected galaxies; \citealt{reddy12b}) is $\beta_0 = -2.52$.
Below, we also consider the more traditionally used \citet{bruzual03}
(BC03) models.

We calculated the IRX-$\beta$ relation assuming an energy balance
between flux that is absorbed and that which is re-emitted in the
infrared \citep{meurer99}.  The absorption is determined by the
extinction or attenuation curve, and we considered several choices
including the SMC extinction curve of \citet{gordon03}, and the
\citet{calzetti00} and \citet{reddy15} attenuation curves.  The
original forms of these extinction/attenuation curves were empirically
calibrated at $\lambda \ga 1200$\,\AA.  The \citet{calzetti00} and
\citet{reddy15} curves were extended down to $\lambda = 950$\,\AA\,
using a large sample of Lyman Break galaxy spectra at $z\sim 3$ and a
newly-developed iterative method presented in \citet{reddy16a}.  The
SMC curve of \citet{gordon03} was extended in the same way, and we
used these extended versions of the curves in this analysis.  For
reference, our new constraints on the shape of dust obscuration curves
imply a lower attenuation of $\lambda\la 1250$\,\AA\, photons relative
to that predicted from polynomial extrapolations below these
wavelengths \citep{reddy16a}.  In practice, because most of the dust
heating arises from photons with $\lambda>1200$\,\AA, the
implementation of the new shapes of extinction/attenuation curves does
little to alter the predicted IRX-$\beta$ relation.  For reference,
the following equations give the relationship between $\ebmv$ and
$\beta$ for the fiducial (BPASS) model with nebular continuum emission
and the shapes of the attenuation/extinction curves derived above:
\begin{eqnarray}
{\bf \rm Calzetti+00:} & \beta =  -2.616 + 4.684\times\ebmv; \nonumber \\
{\bf \rm SMC:} & \beta  =  -2.616 + 11.259\times\ebmv; \nonumber \\
{\bf \rm Reddy+15:} & \beta  =  -2.616 + 4.594\times\ebmv.
\label{eq:betaebmv}
\end{eqnarray}
The intercepts in the above equations are equal to $-2.520$ for the
$300$\,Myr BPASS model.  

For each value of $\ebmv$, we applied the aforementioned dust curves
to the BPASS model and calculated the flux absorbed at $\lambda >
912$\,\AA.  Based on the high covering fraction ($\ga 92\%$) of
optically-thick $\hi$ inferred for $z\sim 3$ galaxies
\citep{reddy16b}, we assumed a zero escape fraction of ionizing
photons and that photoelectric absorption dominates the depletion of
such photons, rather than dust attenuation \citep{reddy16b}.  We then
calculated the resultant Ly$\alpha$ flux assuming Case B recombination
and the amount of Ly$\alpha$ flux absorbed given the value of the
extinction/attenuation curve at $\lambda=1216$\,\AA, and added this to
the absorbed flux at $\lambda>912$\,\AA.  This total absorbed flux is
equated to $L_{\rm IR}$, where we have assumed that all of the dust
emission occurs between $8$ and $1000$\,$\mu$m.  

Finally, we divided the infrared luminosity by the UV luminosity of
the reddened model at $1600$\,\AA\, to arrive at the value of IRX.
The UV slope was computed directly from the reddened model using the
full set of \citet{calzetti94} wavelength windows.  Below, we also
consider the value of $\beta$ computed using the subset of the
\citet{calzetti94} windows at $\lambda < 1740$\,\AA, as well as a
single window spanning the range $1300-1800$\,\AA.  Formally, we find
the following relations between IRX and $\beta$ given
Equation~\ref{eq:betaebmv}, where $\beta$ is measured using the full
set of \citet{calzetti94} wavelength windows:
\begin{eqnarray}
{\bf \rm Calzetti+00:} & {\rm IRX} = 1.67 \times [10^{0.4(2.13\beta + 5.57)} - 1]; \nonumber \\
{\bf \rm SMC:} & {\rm IRX} = 1.79 \times [10^{0.4(1.07\beta + 2.79)} - 1]; \nonumber \\
{\bf \rm Reddy+15:} & {\rm IRX} = 1.68 \times [10^{0.4(1.82\beta + 4.77)} - 1].
\end{eqnarray}
These relations may be shifted redder by $\delta\beta = 0.096$ to
reproduce the IRX-$\beta$ relations for the 300\,Myr BPASS model.  For
reference, Appendix~\ref{sec:sumrelations} summarizes the relations
between $\beta$ and $\ebmv$ and between IRX and $\beta$ for different
assumptions of the stellar population model, nebular continuum,
Ly$\alpha$ heating, and the normalization of the dust curve.

\begin{figure*}
\epsscale{1.0}
\plotone{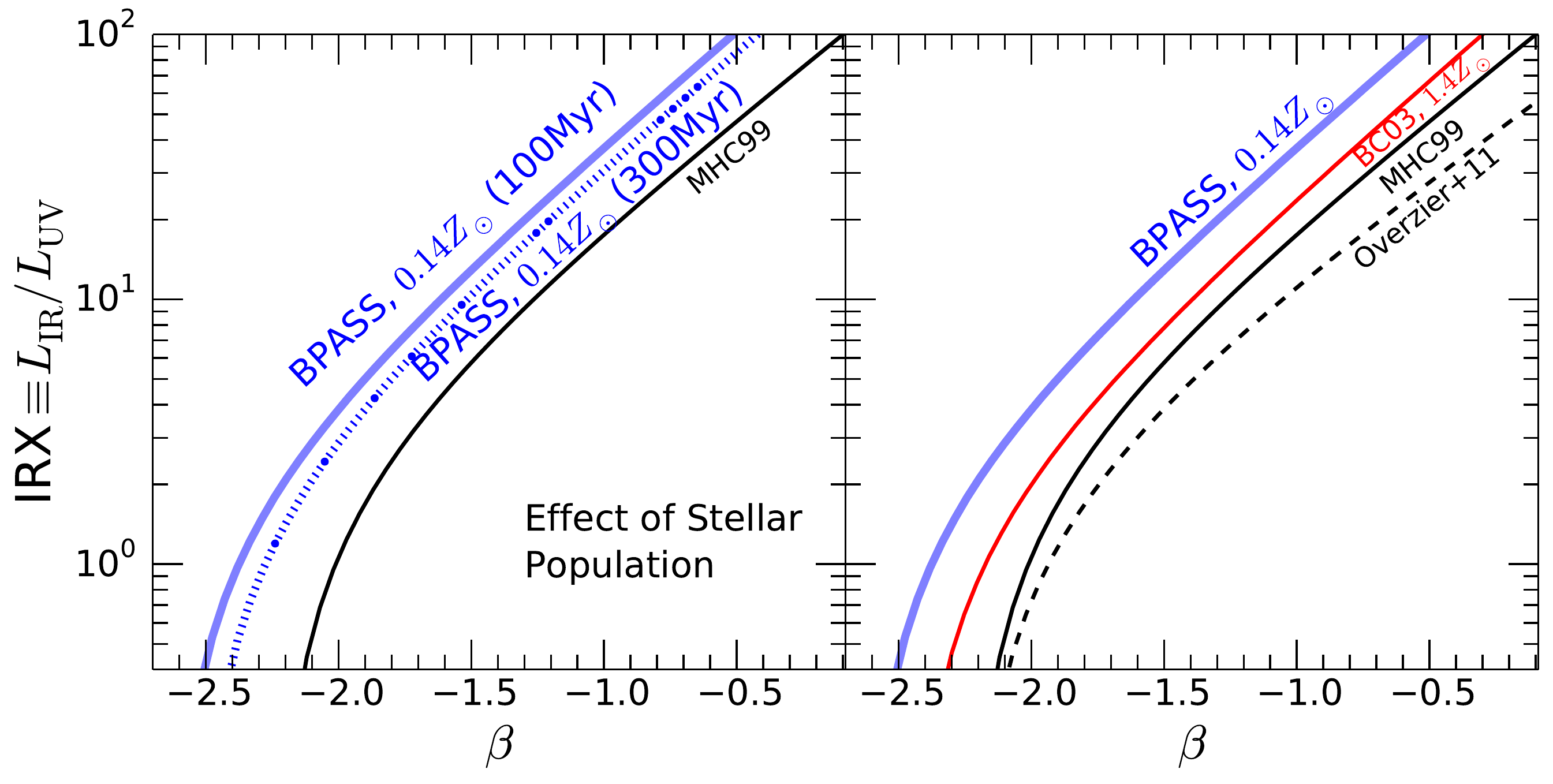}
\caption{Predicted IRX-$\beta$ relations for different assumptions of
  the stellar population.  {\em Left:} IRX-$\beta$ relation for the
  fiducial $0.14Z_\odot$ BPASS model with constant star formation and
  an age of 100\,Myr assuming the \citet{calzetti00} dust attenuation
  curve.  The solid black line shows the \citet{meurer99} relation,
  shifted $0.24$\,dex upward to account for the flux difference
  between the far-infrared ($40-120$\,$\mu$m) passband used in that
  study and the total infrared ($8-1000$\,$\mu$m) passband assumed
  here.  The dotted line indicates the $0.14Z_\odot$ BPASS model with
  an age of 300\,Myr.  {\em Right:} Comparison of our fiducial BPASS
  model with the ``solar metallicity'' \citet{bruzual03} model which, given
  the currently measured solar abundances \citep{asplund09}, equates
  to $1.4Z_\odot$.  The latter also assumes a constant star formation
  with an age of 100\,Myr.  Also shown are the \citet{meurer99} curve
  and the update to this curve provided in \citet{overzier11}.  A
  $0.28Z_\odot$ \citet{bruzual03} model results in an IRX-$\beta$
  relation which is essentially identical to that of the BPASS model
  for the same age.  The shifts in the IRX-$\beta$ relations between
  the models are attributable primarily to differences in the
  intrinsic UV slope, with even the commonly-used \citet{bruzual03}
  model having $\beta_0 = -2.44$ (without including nebular
  continuum), substantially bluer than the $\beta_0 =-2.23$ model
  adopted in \citet{meurer99}.}
\label{fig:irxpred1}
\end{figure*}

\begin{figure*}
\epsscale{1.00}
\plotone{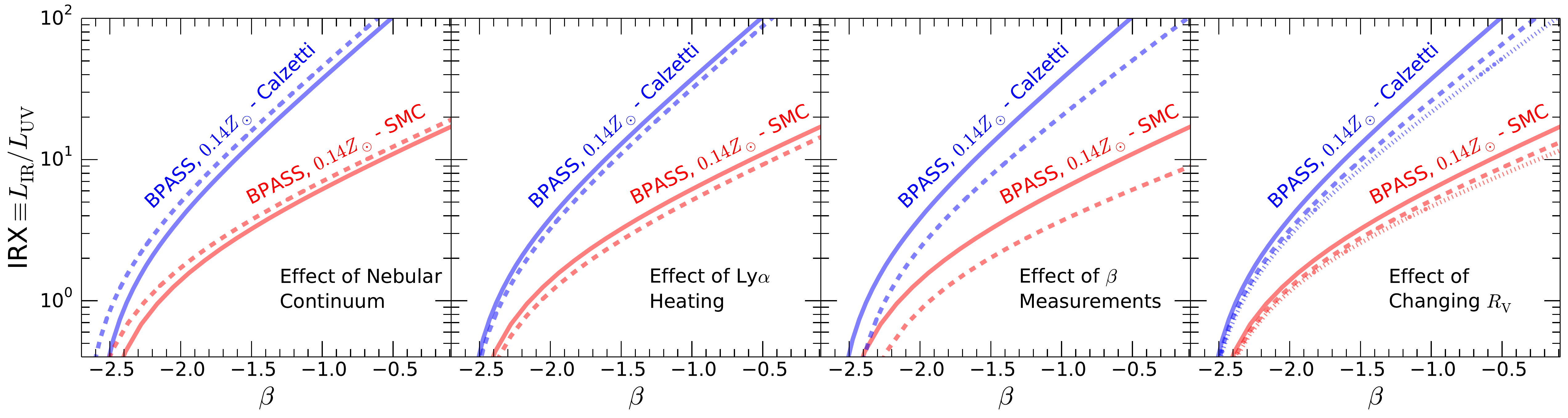}
\caption{Predicted IRX-$\beta$ relations for different assumptions of
  the contribution of nebular continuum emission, effect of Ly$\alpha$
  heating, systematics associated with the measurement of the UV
  slope, and the normalization of the dust curves.  {\em Left:}
  IRX-$\beta$ relations for the fiducial $0.14Z_\odot$ BPASS model,
  with (solid line) and without (dashed line) including nebular
  continuum emission.  Relations are show for both the
  \citet{calzetti00} and SMC dust curves.  Neglecting the contribution
  to the SED from nebular continuum emission will cause one to measure
  a slightly bluer UV slope ($\delta\beta \simeq 0.1$).  {\em Middle
    Left:} IRX-$\beta$ relations for the fiducial model, assuming the
  \citet{calzetti00} attenuation and the SMC extinction curves (solid
  lines).  The dashed lines indicate the result if we assume that none
  of the Ly$\alpha$ emission from the galaxy is absorbed by dust.
  {\em Middle Right:} Same as the middle left panel, where the dashed
  lines now indicate the IRX-$\beta$ relations if $\beta$ is measured
  using a window spanning the range $1300-1800$\,\AA.  {\em Right:}
  Same as the middle left panel, where the dashed and dotted lines now
  show the effect of lowering the total-to-selective extinction by
  $\delta R_{\rm V} = 1.0$ and $1.5$, respectively.}
\label{fig:irxpred2}
\end{figure*}

Figures~\ref{fig:irxpred1} and \ref{fig:irxpred2} convey 
a sense for how the stellar population and nebular continuum,
Ly$\alpha$ heating, UV slope measurements, and the total-to-selective
extinction ($R_{\rm V}$) affect the IRX-$\beta$ relation.  Models with
a bluer intrinsic UV slope require a larger degree of dust obscuration
to reproduce a given observed UV slope, thus causing the IRX-$\beta$
relation to shift towards bluer $\beta$.  Relative to the
\citet{meurer99} relation, the IRX-$\beta$ relations for the fiducial
(BPASS) $100$ and $300$\,Myr models predict a factor of $\approx 2$
more dust obscuration at a given $\beta$ for $\beta \ga -1.7$, and an
even larger factor for $\beta$ bluer than this limit (left panel of
Figure~\ref{fig:irxpred1}).  The commonly utilized BC03 model results
in a factor of $\approx 30\%$ increase in the IRX at a given $\beta$
relative to the \citet{meurer99} curve, while the $0.28Z_\odot$ BC03
model results in an IRX-$\beta$ relation that is indistinguishable
from that of the BPASS model for the same age (right panel of
Figure~\ref{fig:irxpred2}).  These predictions underscore the
importance of the adopted stellar population model when using the
IRX-$\beta$ relation to discern between different dust
attenuation/extinction curves (e.g., \citealt{meurer99, boquien12,
  schaerer13}).  Note that the inclusion of nebular continuum emission
shifts the IRX-$\beta$ relation by $\delta\beta \simeq 0.1$ to the
right (i.e., making $\beta$ redder), so that the IRX at a given
$\beta$ is $\approx 0.1$\,dex lower (leftmost panel of
Figure~\ref{fig:irxpred2}).

The specific treatment of dust heating from Ly$\alpha$ photons has a
much less pronounced effect on the IRX-$\beta$ relation. If none of
the Ly$\alpha$ flux is absorbed by dust---also equivalent to assuming
that the escape fraction of ionizing photons is $100\%$---then the
resulting IRX is $\approx 10\%$ lower at a given $\beta$ than that
predicted by our fiducial model.  Similarly, assuming that all of the
Ly$\alpha$ is absorbed by dust results in an IRX-$\beta$ relation that
is indistinguishable from that of the fiducial model.

\begin{figure*}
\epsscale{1.00}
\plotone{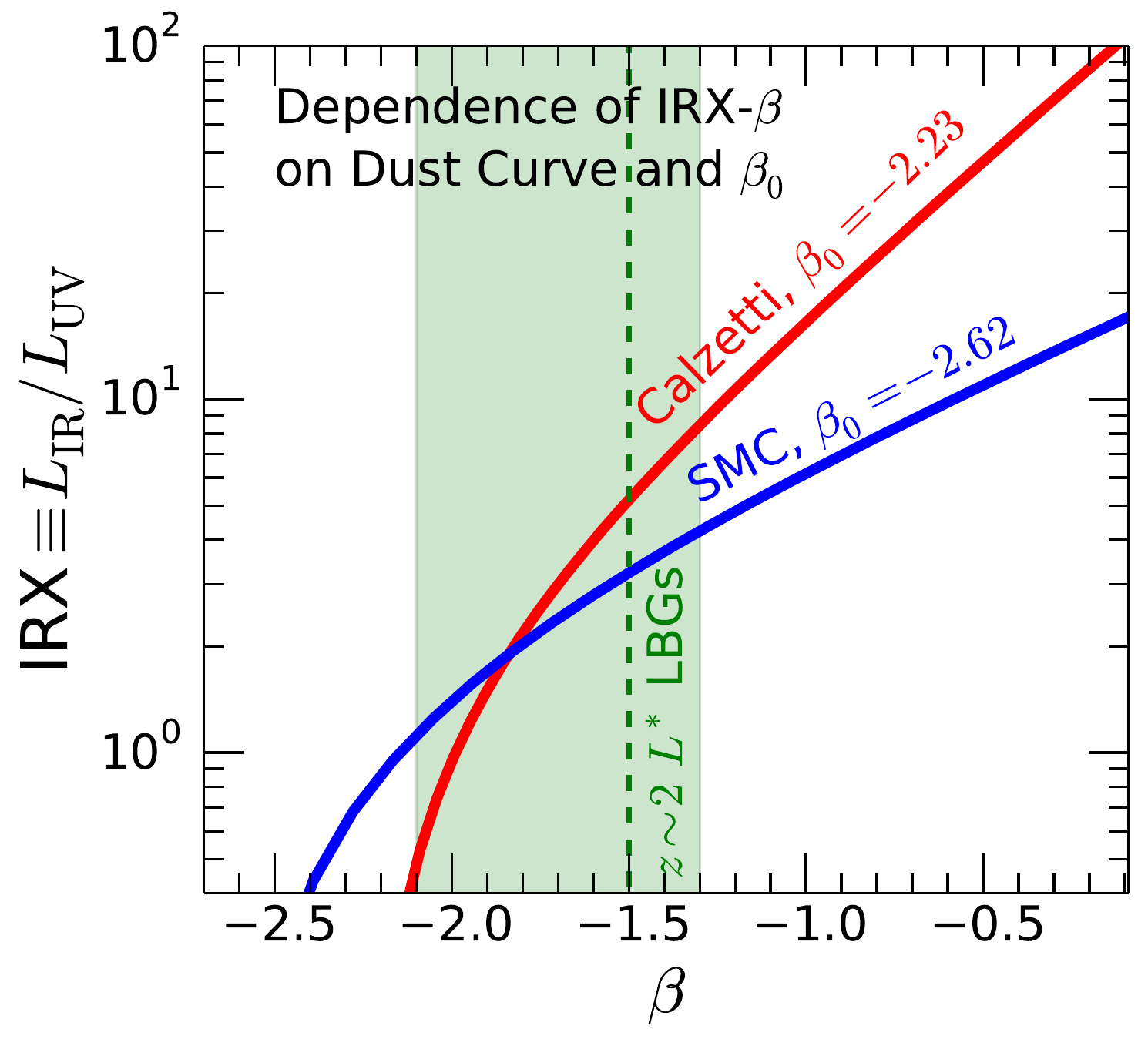}
\caption{Predicted IRX-$\beta$ relation for our fiducial model,
  shifted to have an intrinsic UV slope of $\beta_0 = -2.23$, assuming
  the \citet{calzetti00} dust attenuation curve (red line).  The blue
  curve shows the fiducial model when assuming $\beta_0 = -2.62$ and
  the SMC curve.  There is a substantial overlap (within an factor of
  two in IRX) between the two IRX-$\beta$ relations over the range
  $-2.1\la \beta\la -1.3$ (light green shaded region), where
  $L^{\ast}$ LBGs at $z\sim 2-3$ tend to lie (vertical dashed line;
  \citealt{reddy12a}).}
\label{fig:keyplot}
\end{figure*}

The wavelengths over which $\beta$ is computed will also effect the
IRX-$\beta$ relation to varying degrees, depending on the specific
wavelength ranges and the stellar population model.  For the BPASS
model, computing $\beta$ from the reddened model spectrum within a
single window spanning the range $1300-1800$\,\AA\, results in an
IRX-$\beta$ relation that is shifted by as much as $\delta\beta = 0.4$
to redder slopes.  This effect is due to the fact that the stellar
continuum rises less steeply towards shorter wavelengths for $\lambda
\la 1500$\,\AA.  Consequently, the $\log({\rm IRX})$ is $\simeq
0.18$\,dex lower in this case relative to that computed based on the
full set of \citet{calzetti94} windows.  Similar offsets are observed
when using the subset of the \citet{calzetti94} windows lying at
$\lambda < 1800$\,\AA, while the offsets are not as large with the
BC03 model.  Most previous studies of the IRX-$\beta$ relation adopted a
$\beta$ computed over relatively broad wavelength ranges coinciding
with the \citet{calzetti94} windows.  However, the systematic offsets
in the IRX-$\beta$ relation arising from the narrower wavelength range
used to compute UV slopes become relevant for very high-redshift
(e.g., $z\ga 8$) galaxies where {\em Hubble} photometry is typically
used to constrain the UV slope and where such observations only go as
red as rest-frame $\la 1800$\,\AA.

Finally, the rightmost panel of Figure~\ref{fig:irxpred2} shows the
effect of lowering the total-to-selective extinction ($R_{\rm V}$), or
normalization, of the attenuation/extinction curves by various
amounts.

Of the {\em physical} factors discussed above, the IRX-$\beta$
relation is most sensitive to the effects of changing the intrinsic UV
slope and/or $R_{\rm V}$.  To underscore the importance of the assumed
stellar population when interpreting the IRX-$\beta$ relation, we show
in Figure~\ref{fig:keyplot} the comparison of our fiducial BPASS model
assuming the \citet{calzetti00} curve and an intrinsic $\beta_0 =
-2.23$ (accomplished by simply shifting the model to asymptote to this
intrinsic value), along with the same model assuming an SMC curve with
$\beta_0 = -2.62$.  As is evident from this figure, the two
IRX-$\beta$ relations that assume different attenuation curves and
intrinsic UV slopes have a significant overlap (within a factor two in
IRX) over the range $-2.1\la \beta \la -1.3$.  Notably this range
includes the typical $\beta \simeq -1.5$ found for UV-selected
galaxies at $z\sim 2$ (see \citealt{reddy12a}).  In the next section,
we examine these effects further in the context of the stacked
constraints on IRX-$\beta$ provided by the combined HDUV and 3D-HST
samples.

\section{
DISCUSSION}
\label{sec:discussion}

\subsection{IRX-$\beta$ for the Entire Sample}
\label{sec:irxbeta}

\begin{figure*}
\epsscale{1.1}
\plotone{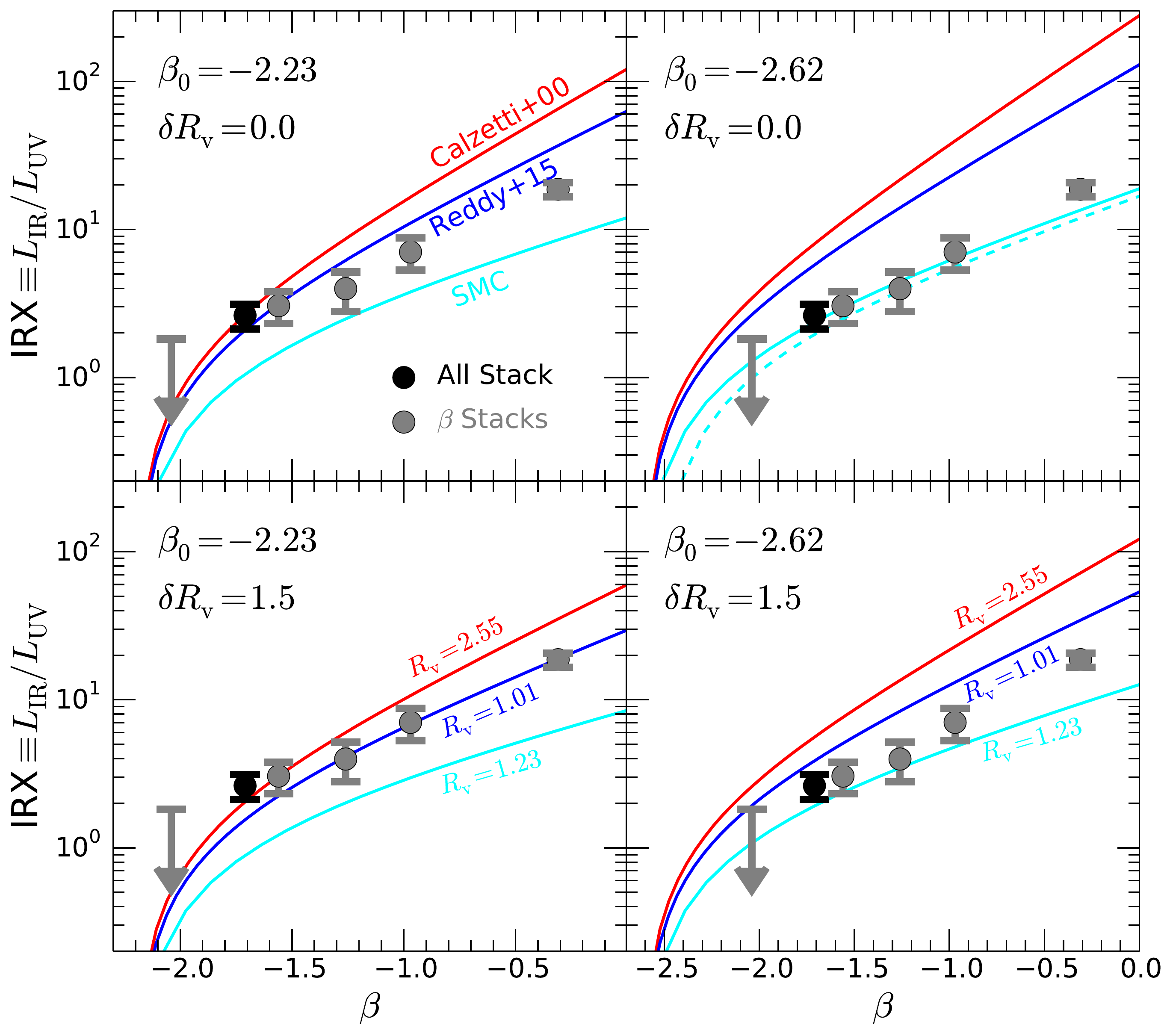}
\caption{Predicted IRX-$\beta$ relations assuming the
  \citet{calzetti00}, \citet{reddy15}, and SMC dust curves---the
  \citet{overzier11} IRX-$\beta$ relation is indistinguishable from
  that obtained from \citet{reddy15} attenuation curve---for the
  fiducial stellar population model (BPASS).  The intrinsic slope of
  this model is $\beta_0 = -2.62$.  In the left-hand panels, the
  predicted IRX-$\beta$ relations have been shifted to show the effect
  of assuming a redder intrinsic UV slope of $\beta_0 = -2.23$, the
  same as that in \citet{meurer99}. The two bottom panels show the
  effect of lowering the normalization of the dust curves by $\delta
  R_{\rm V} = 1.5$, with the specific values of $R_{\rm v}$ indicated.
  The gray points in each panel denote our stacked measurements of IRX
  for galaxies in bins of $\beta$, while the black points shows the
  result of the stack for all galaxies.  The dashed line in the upper
  right-hand panel indicates the IRX-$\beta$ relation implied by the
  SMC extinction curve for an age of 300\,Myr.}
\label{fig:irxbetarv}
\end{figure*}

As a first step in constraining the IRX-$\beta$ relation at
$z=1.5-2.5$, we stacked galaxies in bins of UV slope.  The resulting
IRX for each of these bins, as well as for the whole sample, are shown
in Figure~\ref{fig:irxbetarv}.  The predicted IRX-$\beta$ relations
for different assumptions of the stellar population (BPASS or BC03)
intrinsic UV slope, $\beta_0$, and the difference in normalization of
the dust curves, $\delta R_{\rm V}$, are also shown.  To account for
the former, we simply shifted the fiducial relation (computed assuming
$\beta_0 = -2.62$) so that it asymptotes to a redder value of $\beta_0
= -2.23$, similar to that assumed in \citet{meurer99}.

\begin{figure*}
\epsscale{1.0}
\plotone{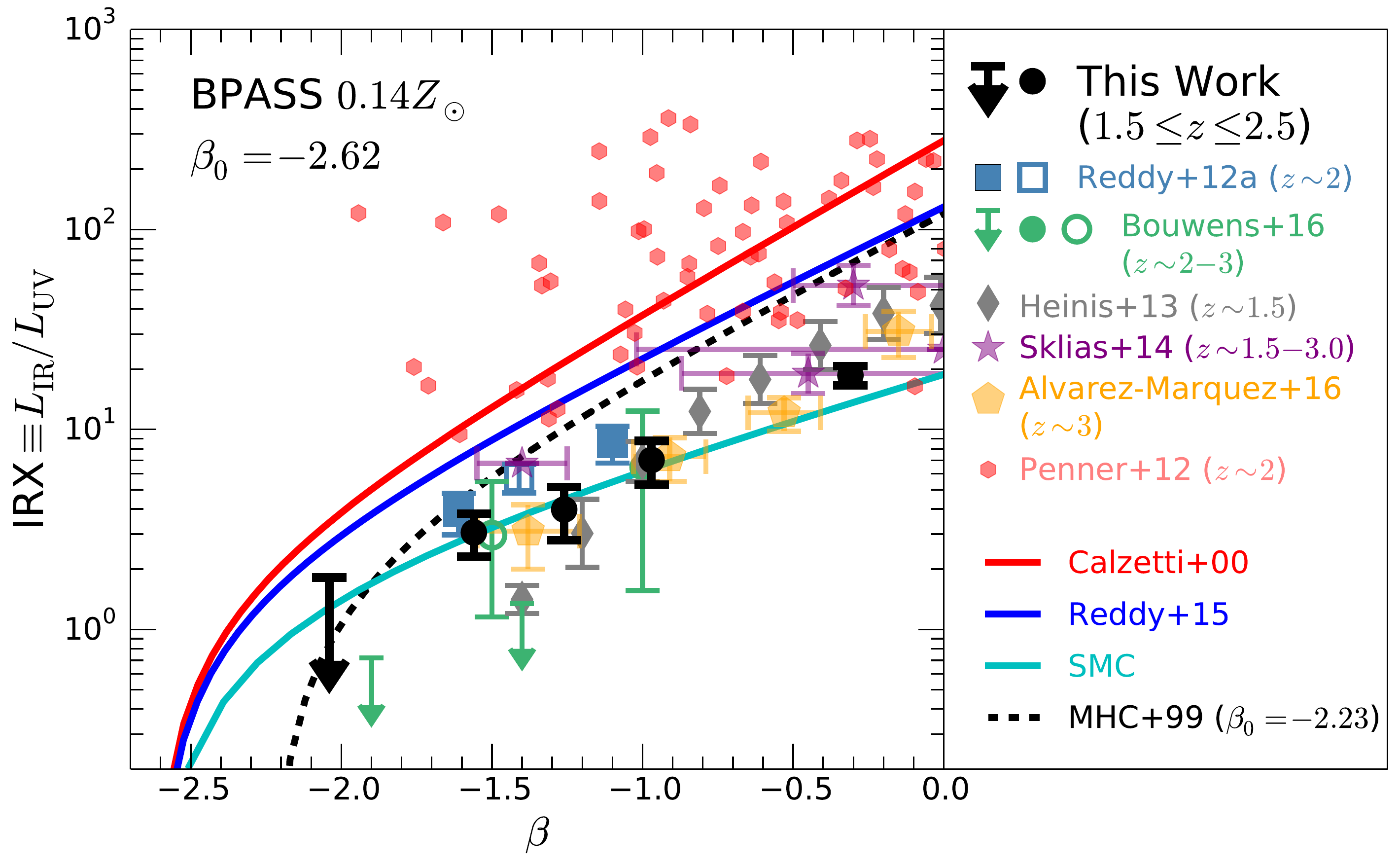}
\caption{Comparison of our IRX-$\beta$ measurements with several from
  the literature for (primarily) UV-selected galaxies at $1.5\la z\la
  3.0$, including those from \citet{reddy12a}, \citet{bouwens16b},
  \citet{heinis13}, and \citet{alvarez16}.  Measurements are also
  shown for small samples of gravitationally-lensed and dust obscured
  galaxies from \citet{sklias14} and \citet{penner12}, respectively.
  The predicted IRX-$\beta$ relations (see Section~\ref{sec:predirx})
  for our fiducial model and the SMC, \citet{reddy15}, and
  \citet{calzetti00} dust curves are indicated, along with the
  original \citet{meurer99} relation which assumed $\beta_0 = -2.23$.
}
\label{fig:irxbetalit}
\end{figure*}

Our stacked results indicate a highly significant ($\ga 20\sigma$)
correlation between IRX and $\beta$.  However, none of the predicted
relations calculated based on assuming an intrinsic UV slope of
$\beta_0 = -2.23$, as in \citet{meurer99}, are able to reproduce our
stacked estimates for the full range of $\beta$ considered here.  For
example, the upper left panel of Figure~\ref{fig:irxbetarv} shows that
while both the \citet{calzetti00} and \citet{reddy15} attenuation
curves predict IRX that are within $3\sigma$ of our stacked values for
$\beta < -1.2$, they over-predict the IRX for galaxies with redder
$\beta$.  

Lowering the normalization of the \citet{reddy15} attenuation curve by
$\delta R_{\rm V} = 1.5$ results in a better match to the stacked
determinations, but with some disagreement (at the $>3\sigma$ level)
with the stack of the entire sample (lower left panel of
Figure~\ref{fig:irxbetarv}).  \citet{reddy15} estimated the systematic
uncertainty in their determination of $R_{\rm V}$ to be $\delta R_{\rm
  V}\approx 0.4$, which suggests that their curve may not have a
normalization as low as $R_{\rm V} = 1.0$ given their favored value of
$R_{\rm V} = 2.51$. Regardless, without any modifications to the
normalizations and/or shapes of the attenuation curves in the
literature \citep{calzetti00, gordon03, reddy15}, the corresponding
IRX-$\beta$ relations are unable to reproduce our stacked estimates if
we assume an intrinsic UV slope of $\beta_0 = -2.23$.  At face value,
these results suggest that the attenuation curve describing our sample
is steeper than the typically utilized \citet{calzetti00} relation,
but grayer than the SMC extinction curve.  However, this conclusion
depends on the intrinsic UV slope of the stellar population, as we
discuss next.

Independent evidence favors the low-metallicity BPASS model in
describing the underlying stellar populations of $z\sim 2$ galaxies
\citep{steidel16}.  The very blue intrinsic UV slope characteristic of
this model---as well as those of the BC03 models with comparable
stellar metallicities (e.g., the $0.28Z_\odot$ BC03 model with the
same high-mass power-law index of the IMF as the BPASS model has
$\beta_0 = -2.65$)---is also favored in light of the non-negligible
number of galaxies in our sample ($\approx 9\%$) that have
$\beta<-2.23$ at the $3\sigma$ level, the canonical value assumed in
\citet{meurer99}.  Figure~\ref{fig:irxpred1} shows that the
low-metallicity models with blue $\beta_0$ result in IRX-$\beta$
relations that are significantly shifted relative to those assuming
redder $\beta_0$.  With such models, we find that our stacked
measurements are best reproduced by an SMC-like extinction curve
(upper right-hand panel of Figure~\ref{fig:irxbetarv}), in the sense
that all of the measurements lie within $3\sigma$ of the associated
prediction.  On the other hand, with such stellar population models,
grayer attenuation curves (e.g., \citealt{calzetti00}) over-predict
the IRX at a given $\beta$ by a factor of $\approx 2-7$.  More
generally, we find that the slope of the IRX-$\beta$ relation implied
by our stacked measurements is better fit with that obtained when
considering the SMC extinction curve, while grayer attenuation curves
lead to a more rapid rise in IRX with increasing $\beta$.

Our stacked measurements and predicted IRX-$\beta$ curves are compared
with several results from the literature in
Figure~\ref{fig:irxbetalit}.  In the context of the IRX-$\beta$
predictions that adopt sub-solar metallicities, we find that most of
the stacked measurements for UV-selected galaxies at $z\sim 1.5-3.0$
suggest a curve that is SMC-like, at least for $\beta \la -0.5$.
Several of the samples, including those of \citet{heinis13},
\citet{alvarez16}, and \citet{sklias14}, indicate an IRX that is
larger than the SMC prediction for $\beta \ga -0.5$.  Such a behavior
is not surprising given that the dust obscuration has been shown to
decouple from the UV slope for galaxies with large star-formation
rates, as is predominantly the case for most star-forming galaxies
with very red $\beta$ \citep{goldader02, chapman05, reddy06a, reddy10,
  penner12, casey14b, salmon16}.  

As discussed in a number of studies \citep{reddy06a, reddy10,
  penner12, casey14b, koprowski16}, dusty galaxies in general can
exhibit a wide range in $\beta$ (from very blue to very red) depending
on the particular spatial configuration of the dust and UV-emitting
stars.  Figure~\ref{fig:irxbetalit} shows that the degree to which
such galaxies diverge from a given attenuation curve depends on
$\beta_0$.  Many of the dusty galaxies that would appear to have IRX
larger than the \citet{meurer99} or \citet{calzetti00} predictions may
in fact be adequately described by such curves if the stellar
populations of these galaxies are characterized by very blue intrinsic
UV spectral slopes.  On the other hand, if these dusty galaxies have
relatively enriched stellar populations, and redder intrinsic slopes,
then their departure from the \citet{calzetti00} prediction would be
enhanced.

\begin{figure*}
\epsscale{1.0}
\plotone{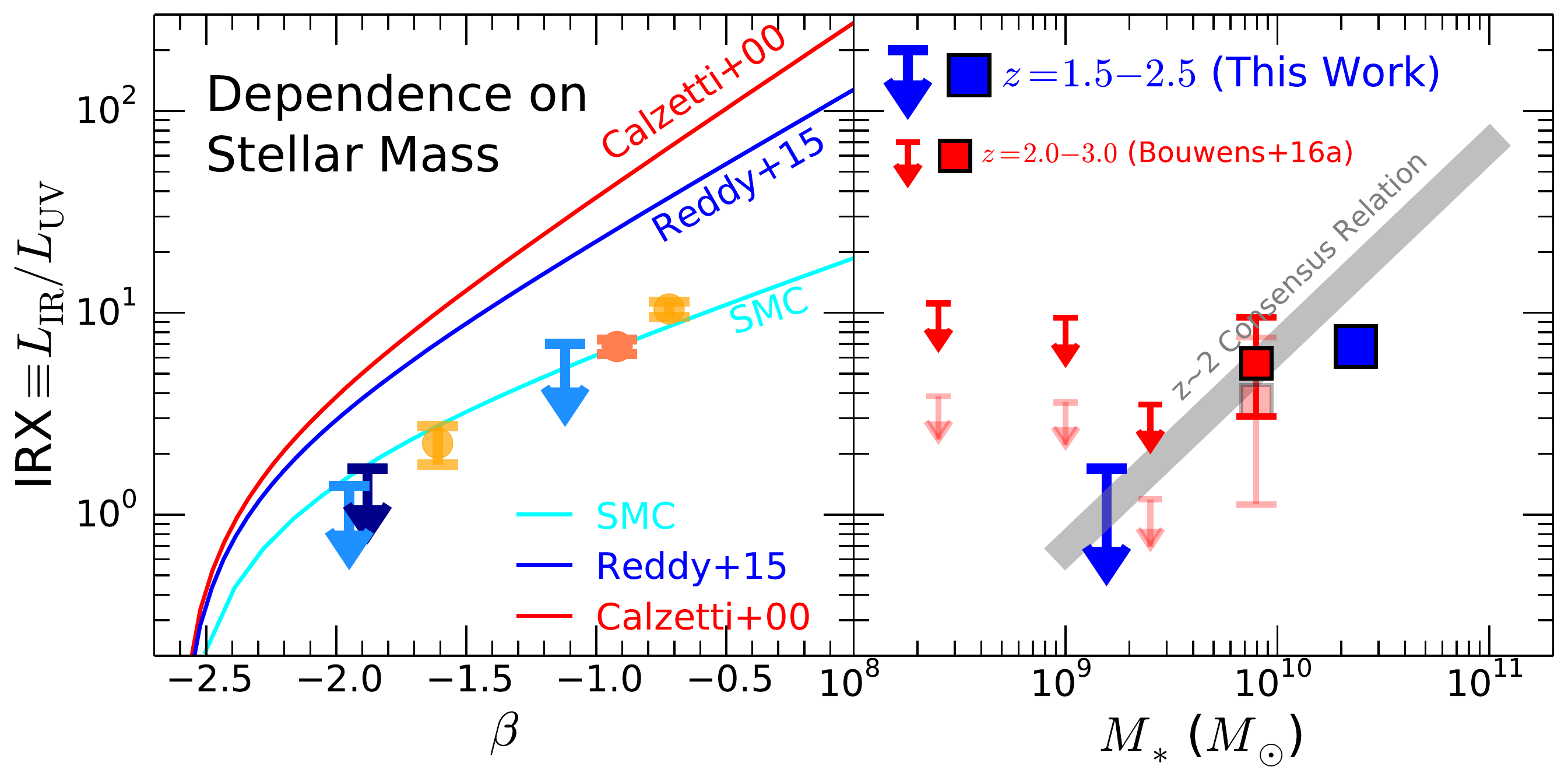}
\caption{{\em Left:} IRX-$\beta$ relation as a function of stellar
  mass.  The dark blue arrow indicates the $3\sigma$ upper limit for
  low-mass galaxies with $M_{\ast}\le 10^{9.75}$\,$M_\odot$, while the
  light blue arrows are for the low-mass subsample with $\beta\le
  -1.4$ and $\beta > -1.4$.  Similarly, the dark red and orange
  circles indicate the same for the high-mass ($M_{\ast}>
  10^{9.75}$\,$M_\odot$) sample.  Our measurements suggest that
  high-mass galaxies at $z=1.5-2.5$ are consistent with SMC
  extinction, while low-mass galaxies have upper limits in IRX that
  suggest a dust curve that is steeper than the SMC curve.  {\rm
    Right:} IRX as a function of stellar mass for galaxies in our
  sample (blue symbols) and that of \citet{bouwens16b} (red symbols),
  where the arrows indicate $3\sigma$ upper limits.  As the
  \citet{bouwens16b} measurements are based on ALMA 2\,mm continuum
  data, the derived $L_{\rm IR}$ are sensitive to the assumed dust
  temperature.  The dark and light red symbols indicate their results
  when assuming an evolving dust temperature and a constant
  temperature with $T=35$\,K, respectively.  The thick gray line
  denotes the ``consensus relation'' at $z\sim 2$ computed by
  \citet{bouwens16b} and based on data from \citet{reddy10},
  \citet{whitaker14} and \citet{alvarez16}.}
\label{fig:irxmass}
\end{figure*}

Undoubtedly, large variations in IRX can also be expected with
different geometries of dust and stars.  Regardless, if sub-solar
metallicity models are widely representative of the stellar
populations of typical star-forming galaxies at $z\ga 1.5$, then our
stacked measurements, along with those in the literature, tend to
disfavor gray attenuation curves for these galaxies.  The large sample
studied here, as well as those of \citet{bouwens16b} and
\citet{alvarez16}, suggest an SMC-like curve.  At first glance, this
conclusion may appear to be at odds with the wide number of previous
investigations that have found that the \citet{meurer99} and
\citet{calzetti00} attenuation curves generally apply to
moderately-reddened star-forming galaxies at $z\ga 1.5$ (e.g.,
\citealt{nandra02, seibert02, reddy04, reddy06a, daddi07a, pannella09,
  reddy10, magdis10, overzier11, reddy12a, forrest16, debarros16};
c.f., \citealt{heinis13}).  

In the framework of our present analysis, the reconciliation between
these results is simple.  Namely, our analysis does not call into
question previous {\em measurements} of IRX-$\beta$, but calls for a
different {\em interpretation} of these measurements.  In the previous
interpretation, most of the stacked measurements from the literature
were found to generally agree with the \citet{meurer99} relation {\em
  if we assume a relatively red intrinsic slope of $\beta=-2.23$}.  In
the present interpretation, we argue that sub-solar metallicity models
necessitate a steeper attenuation curve in order to reproduce the
measurements of IRX-$\beta$ (e.g., see also \citealt{cullen17}).  Our
conclusion is aided by the larger dynamic range in $\beta$ probed by
the HDUV and 3D-HST samples, which allows us to better discriminate
between different curves given that their corresponding IRX-$\beta$
relations separate significantly at redder $\beta$
(Figures~\ref{fig:keyplot} and \ref{fig:irxbetalit}).

\subsection{IRX Versus Stellar Mass}
\label{sec:irxmass}

The well-studied correlations between star-formation rate and stellar
mass (e.g., \citealt{noeske07, reddy06a, daddi07a, pannella09,
  wuyts11, reddy12b, whitaker14, schreiber15, shivaei15}), and between
star-formation rate and dust attenuation (e.g., \citealt{wang96,
  adelberger00, reddy06b, buat07, buat09, burgarella09, reddy10}),
have motivated the use of stellar mass as a proxy for attenuation
\citep{pannella09, reddy10, whitaker14, pannella15, alvarez16,
  bouwens16b} as the former can be easily inferred from fitting stellar
population models to broadband photometry.  The connection between
reddening and stellar mass can also be deduced from the
mass-metallicity relation \citep{tremonti04, kewley08, andrews13,
  erb06a, maiolino08, henry13, maseda14, steidel14, sanders15}.

Motivated by these results, we stacked galaxies in two bins of stellar
mass divided at $M_{\ast} = 10^{9.75}$\,$M_\odot$ (and further
subdivided into bins of $\beta$; Table~\ref{tab:stackedresults}) to
investigate the dependence of the IRX-$\beta$ relation on stellar
mass.\footnote{The stellar masses obtained with \citet{conroy10}
  models (Section~\ref{sec:sample}) are on average within $0.1$\,dex
  of those obtained assuming the fiducial (BPASS) model with the same
  \citep{chabrier03} IMF.}  The high-mass subsample ($M_\ast >
10^{9.75}$\,$M_\odot$) exhibits a redder UV slope
($\langle\beta\rangle = -0.92$) and larger IRX ($\langle{\rm
  IRX}\rangle=6.8\pm 0.6$) than the low-mass subsample with
$\langle\beta\rangle = -1.88$ and $\langle{\rm IRX}\rangle<1.7$
($3\sigma$ upper limit).  Moreover, the high-mass subsample exhibits
an IRX-$\beta$ relation consistent with that predicted assuming our
fiducial stellar population model and the SMC extinction curve
(Figure~\ref{fig:irxmass}).  Separately, the low-mass subsample as a
whole, as well as the subset of the low-mass galaxies with $\beta \le
-1.4$, have $3\sigma$ upper limits on IRX that require a dust curve
that is at least as steep as the SMC.

The constraints on the IRX-$M_{\ast}$ relation from our sample are
shown relative to previous determinations in the right panel of
Figure~\ref{fig:irxmass}.  The ``$z\sim 2$ Consensus Relation''
presented in \citet{bouwens16b} was based on the IRX-$M_{\ast}$ trends
published in \citet{reddy10}, \citet{whitaker14}, and
\citet{alvarez16}.  Formally, our stacked detection for the high-mass
($M_{\ast}>10^{9.75}$\,$M_\odot$) subsample lies $\approx 4\sigma$
below the consensus relation, but is in excellent agreement with the
mean IRX found for galaxies of similar masses ($\simeq 2\times
10^{10}$\,$M_\odot$) in \citet{reddy10}.  The upper limit in IRX for
the low-mass ($M_{\ast}\le 10^{9.75}$\,$M_\odot$) sample is consistent
with the predictions from the consensus relation.  Based on these
comparisons, we conclude that the IRX-$M_{\ast}$ relation from the
present work is in general agreement with previous determinations, and
lends support for previous suggestions that stellar mass may be used
as a rough proxy for dust attenuation in high-redshift star-forming
galaxies (e.g., \citealt{pannella09, reddy10, bouwens16b}).  Moreover,
these comparisons underscore the general agreement between our IRX
measurements (e.g., as a function of $\beta$ and $M_{\ast}$) and those
in the literature, in spite of our different interpretation of these
results in the context of the dust obscuration curve applicable to
high-redshift galaxies (Section~\ref{sec:irxbeta}).

\subsection{IRX Versus UV Luminosity}
\label{sec:irxuvlum}

As alluded to in Section~\ref{sec:intro}, quantifying the dust
attenuation of UV-faint (sub-$L^{\ast}$) galaxies has been a
longstanding focus of the high-redshift community.  While the steep
faint-end slopes of UV luminosity functions at $z\ga 2$ imply that
such galaxies dominate the UV luminosity density at these redshifts,
knowledge of their dust obscuration is required to assess their
contribution the cosmic star-formation-rate density (e.g.,
\citealt{steidel99, adelberger00, bouwens07,reddy08}).  Several
studies have argued that UV-faint galaxies are on average less dusty
than their brighter counterparts \citep{reddy08, bouwens09, bouwens12,
  kurczynski14}.  This inference is based on the fact that the
observed UV luminosity is expected to monotonically correlate with
star-formation rate for galaxies fainter than $L^{\ast}$ (e.g., see
Figure~13 of \citealt{reddy10} and Figure~10 of \citealt{bouwens09})
and that the dustiness is a strong function of star-formation rate
\citep{wang96, adelberger00, reddy06b, buat07, buat09, burgarella09,
  reddy10}.  

While several investigations have shown evidence for a correlation
between IRX and UV luminosity \citep{bouwens09, reddy10, reddy12a},
others point to a roughly constant IRX as a function of UV luminosity
\citep{buat09, heinis13}.  As discussed in these studies, the
different findings may be a result of selection biases, in the sense
that UV-selected samples will tend to miss the dustiest galaxies,
which also have faint observed UV luminosities.  Hence, for purely
UV-selected samples, IRX would be expected to decrease with $L_{\rm
  UV}$.  Alternatively, the rarity of highly dust-obscured galaxies
compared to intrinsically faint galaxies (e.g., as inferred from the
shapes of the UV and IR luminosity functions; \citealt{caputi07,
  reddy08, reddy09, magnelli13}) implies that in a number-weighted
sense, the mean bolometric luminosity should decrease towards fainter
$L_{\rm UV}$.  How this translates to the variation of IRX with
$L_{\rm UV}$ will depend on how quickly the dust can build up in
dynamically-relaxed faint galaxies.  From a physical standpoint, dust
enrichment on timescales much shorter than the dynamical timescale
would suggest a relatively constant IRX as a function of $L_{\rm UV}$.

The HDUV/3D-HST sample presents a unique opportunity to evaluate the
trend between IRX and $L_{\rm UV}$ as the selection is based on
rest-optical criteria.  Consequently, our sample is less sensitive to
the bias against dusty galaxies that are expected to be significant in
UV-selected samples (e.g., \citealt{chapman00, barger00, buat05,
  burgarella05, reddy05, reddy08, casey14a}).  Indeed,
Figure~\ref{fig:zmag} shows that our sample includes a large number of
UV-faint galaxies that are also quite dusty based on their red $\beta
\ga -1.4$---these galaxies, while dusty, still have bolometric
luminosities that are sufficiently faint to be undetected in the PACS
imaging.

\begin{figure*}
\epsscale{1.00}
\plotone{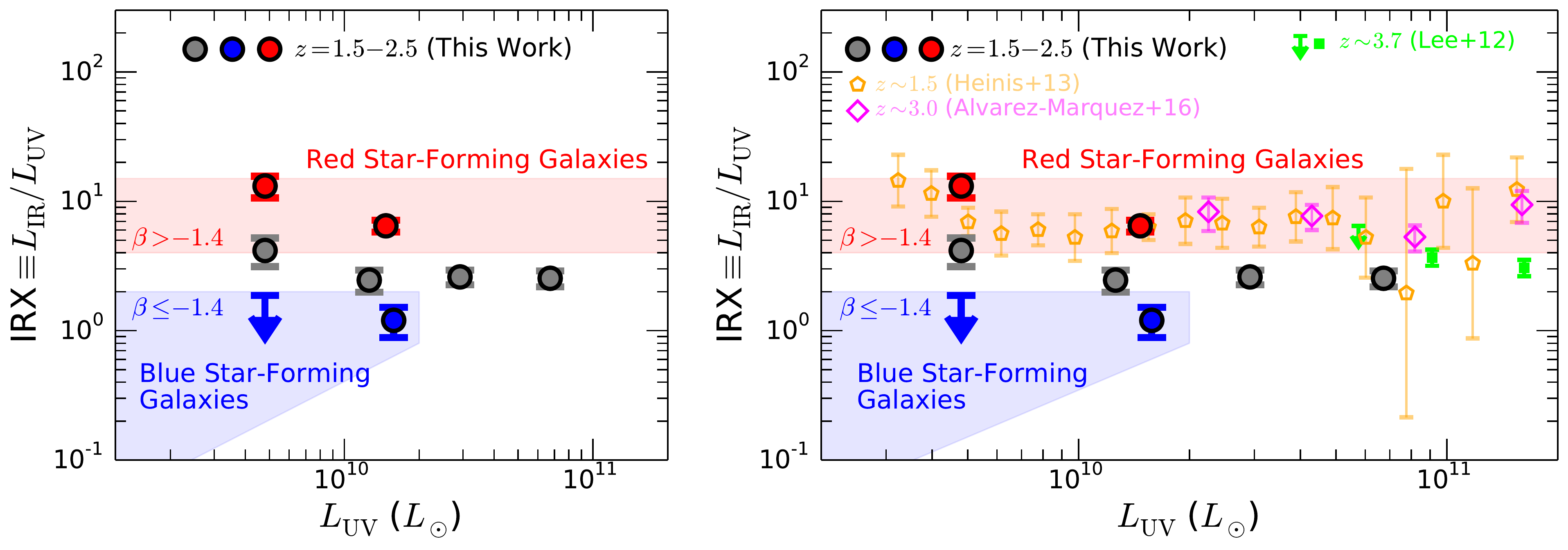}
\caption{Dust attenuation as a function of UV luminosity at $z\sim
  1.5-3.7$.  Our stacked measurements in bins of $L_{\rm UV}$ are
  indicated by the large gray circles, and the large blue and red
  circles indicate the values obtained by further subdividing the
  stacks in bins of $\beta$ (see Table~\ref{tab:stackedresults}).  The
  right panel also shows other results for UV-selected galaxies at
  $z\sim 1.5$ \citep{heinis13}, $z\sim 3$ \citep{alvarez16}, and
  $z\sim 3.7$ \citep{lee12}.  The samples of \citet{heinis13} and
  \citet{alvarez16} include galaxies that are primarily redder ($\beta
  \ga -1.4$) than those in our sample.  When restricting our sample to
  a similar range of $\beta$, we find an IRX-$L_{\rm UV}$ relation
  that is in excellent agreement with that of \citet{heinis13} and
  \citet{alvarez16}.  Unsurprisingly, the IRX-$L_{\rm UV}$ relation
  for blue galaxies with $\beta \le -1.4$ is offset by $\approx
  1$\,dex to lower IRX.}
\label{fig:irxuv}
\end{figure*}

Figure~\ref{fig:irxuv} shows the relationship between dust attenuation
and UV luminosity for our sample (gray points) and those of
\citet{heinis13} at $z\sim 1.5$ and \citet{alvarez16} at $z\sim 3$.
The latter two indicate an IRX that is roughly constant with $L_{\rm
  UV}$, but one which is a factor of $2-3$ offset towards higher IRX
than found for our sample.  This offset is easily explained by the
fact that the IRX-$L_{\rm UV}$ relations for the $z\sim 1.5$ and
$z\sim 3.0$ samples were determined from galaxies that are on average
significantly redder than in our sample.  In particular, most of the
constraints on IRX-$L_{\rm UV}$ from these studies come from galaxies
with $\beta \ga -1.5$.  When limiting our stacks to galaxies with
$\beta > -1.4$ (red points in Figure~\ref{fig:irxuv}), we find an
excellent agreement with the IRX-$L_{\rm UV}$ relations found by
\citet{heinis13} and \citet{alvarez16}.  On the other hand, stacking
those galaxies in our sample with $\beta\le -1.4$ results in an
IRX-$L_{\rm UV}$ relation that is not surprisingly offset towards
lower IRX than for the sample as a whole.

Thus, while the IRX-$L_{\rm UV}$ relation appears to be roughly
constant for all of the samples considered here,
Figure~\ref{fig:irxuv} implies that the $\beta$ distribution as a
function of $L_{\rm UV}$ is at least as important as the presence of
dusty star-forming galaxies in shaping the observed IRX-$L_{\rm UV}$
relation.  Furthermore, the trend of a bluer average $\beta$ with
decreasing $L_{\rm UV}$ (e.g., Figure~\ref{fig:zmag}; see also
\citealt{reddy08, bouwens09}) suggests that the mean reddening should
be correspondingly lower for UV-faint galaxies than for UV-bright ones
once the effect of the less numerous dusty galaxies with red $\beta$
are accounted for.

The blue ($\beta \le -1.4$) star-forming galaxies in our sample have
IRX$\la 1$, such that the infrared and UV luminosities contribute
equally to the bolometric luminosities of these galaxies.  The
expectation of rapid dust enrichment from core-collapse supernovae
\citep{todini01} implies that it is unlikely that the dust obscuration
can be significantly lower than this value when observed for
dynamically-relaxed systems.  Consequently, the observation that the
mean UV slope becomes progressively bluer for fainter galaxies at
high-redshift (e.g., \citealt{bouwens09, wilkins11, finkelstein12,
  alavi14}) may simply be a result of systematic changes in
metallicity and/or star-formation history where the intrinsic UV slope
also becomes bluer but IRX remains relatively constant (e.g.,
Figure~\ref{fig:irxpred1}; \citealt{wilkins11, wilkins13, alavi14}).
{\em Thus, the common observation that UV-faint galaxies are bluer
  than their brighter counterparts may not directly translate into a
  lower dust obscuration for the former.}

Moreover, $\beta$ is relatively insensitive to IRX for
$\beta-\beta_0\la 0.2$ (e.g., Figure~\ref{fig:irxpred1}).  Our results
thus indicate caution when using the IRX-$\beta$ relation to infer the
dust reddening of blue galaxies at high-redshift, as such estimates
may be highly dependent on the intrinsic UV slope of the stellar
population and even otherwise quite uncertain if the difference in
observed and intrinsic UV slopes is small.

\subsection{Young/Low-Mass Galaxies}
\label{sec:young}

ALMA has opened up new avenues for investigating the ISM and dust
content of very high-redshift galaxies, and a few recent efforts have
focused in particular on the [\ion{C}{2}]$158$\,$\mu$m line in
galaxies at $z\ga 5$ \citep{schaerer15, maiolino15, watson15, capak15,
  willott15, knudsen16, pentericci16} and dust continuum at mm
wavelengths.  \citet{capak15} report on ALMA constraints on the IRX of
a small sample of 10 $z\sim 5.5$ LBGs and find that they generally
fall below the SMC extinction curve.  The disparity between the SMC
curve and their data points is increased if one adopts a bluer
intrinsic slope than that assumed in \citet{meurer99}, a reasonable
expectation for these high-redshift and presumably lower metallicity
galaxies.

\begin{figure*}
\epsscale{1.10}
\plotone{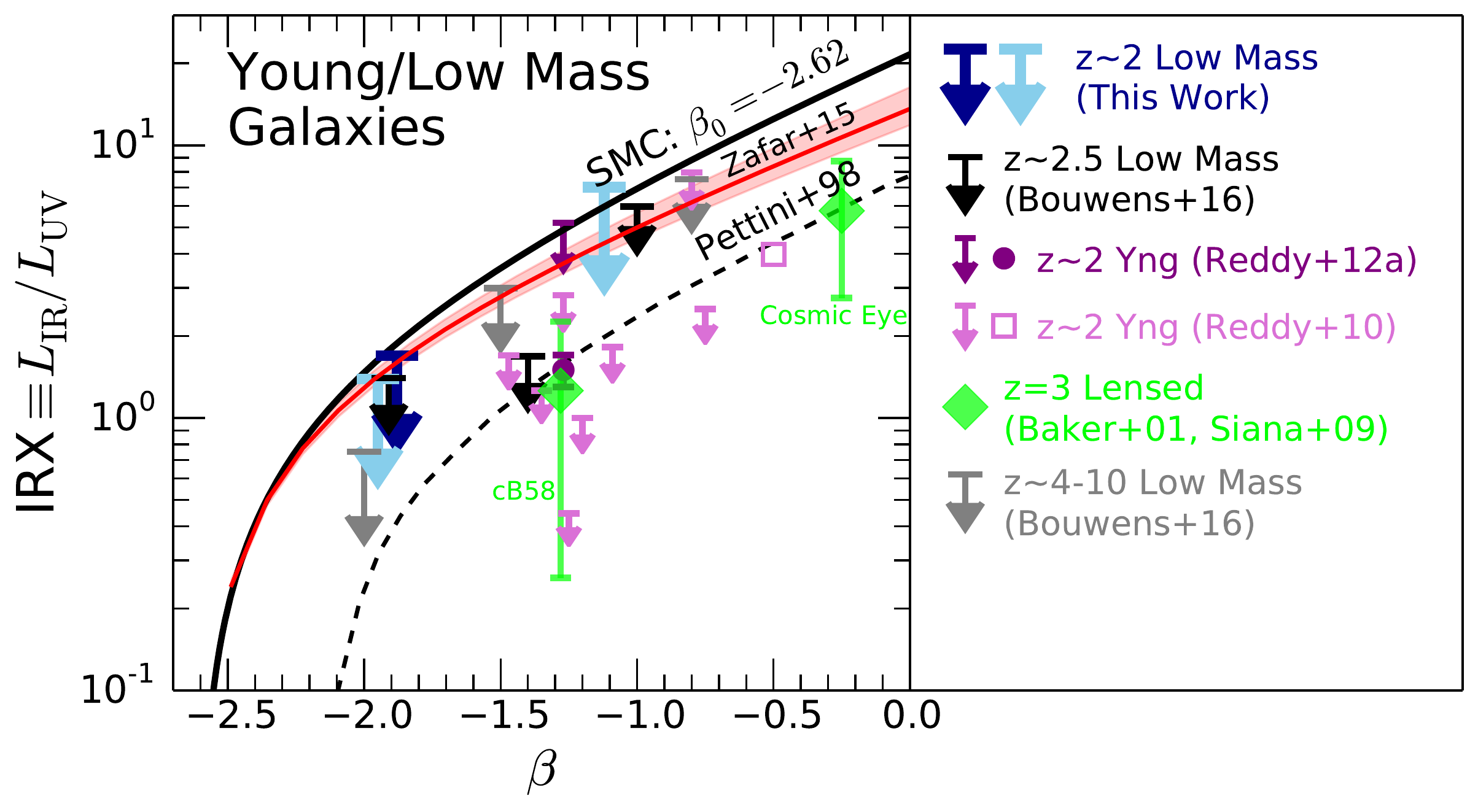}
\caption{IRX-$\beta$ diagram for young and/or low-mass galaxies at
  $z\ga 2$.  The upper limits in IRX for low-mass ($M_{\ast}\le
  10^{9.75}$\,$M_\odot$) galaxies in our sample are indicated by the
  dark and light blue arrows, same as those shown in the left panel of
  Figure~\ref{fig:irxmass}.  Upper limits in IRX derived from ALMA
  measurements for low-mass ($M_{\ast}\le 10^{9.75}$\,$M_\odot$)
  galaxies at $z\sim 2-3$ are indicated by the dark gray arrows
  \citep{bouwens16b}.  The upper limit in IRX derived from a {\em
    Herschel}/PACS stack for UV-selected $z\sim 2$ galaxies with ages
  $\la 100$\,Myr is indicated by the purple arrow, while the purple
  filled circle indicates the stacked inference from {\em
    Spitzer}/MIPS 24\,$\mu$m data \citep{reddy12a}.  Similarly, the
  light purple arrows and open square denote the upper limits for
  individual $<100$\,Myr galaxies and the stacked result,
  respectively, derived from 24\,$\mu$m data \citep{reddy10}.  Also
  shown are results from two lensed galaxies (large green diamonds;
  cB58 and the Cosmic Eye; \citealt{baker01, siana09}), and upper
  limits for low-mass galaxies at $z\sim 4-10$ (light gray arrows;
  \citealt{bouwens16b}).  For reference, the black dashed line shows
  the SMC curve as parameterized by \citet{pettini98}, while the red
  line and shaded region denotes the average and dispersion in the
  IRX-$\beta$ relation for the average extinction curve derived for
  quasars \citep{zafar15}.  For clarity, the results from
  \citet{capak15} for $z\sim 5.5$ LBGs are not shown, but we note that
  most of their points lie at least $1\sigma$ below the
  \citet{pettini98} curve shown in the figure.
The detections and upper limits for young and low-mass galaxies at a
variety of redshifts suggest a dust curve that is on average steeper
than the SMC curve, particularly if we assume a blue intrinsic UV
slope of $\beta_0 = -2.62$ (thick black solid line).  }
\label{fig:irxyoung}
\end{figure*}

More generally, earlier results suggesting that ``young'' LBGs (ages
$\la 100$\,Myr) and/or those with lower stellar masses at $z\ga 2$ are
consistent with an SMC curve \citep{baker01, reddy06a, siana08,
  siana09, reddy10, reddy12a, bouwens16b} would also require a
steeper-than-SMC curve if their intrinsic slopes are substantially
bluer than that normally assumed in interpreting the IRX-$\beta$
relation.  Unfortunately, there is only a very small number of
galaxies in our sample that have SED-determined ages of $<100$\,Myr
(81 galaxies), and stacking them results in an unconstraining upper
limit on IRX (Table~\ref{tab:stackedresults}).  

Note that an ambiguity arises because the ages are derived from
SED-fitting, which assumes some form of the attenuation curve.
Following \citet{reddy10}, the number of galaxies considered ``young''
would be lower under the assumption of SMC extinction rather than
\citet{calzetti00} attenuation, as the former results in lower $\ebmv$
for a given UV slope, translating into older ages.  Self-consistently
modeling the SEDs based on the location of galaxies in the IRX-$\beta$
plane results in fewer $<100$\,Myr galaxies, but of course their
location in the IRX-$\beta$ plane is unaffected \citep{reddy10}, as is
the conclusion that such young galaxies would require a dust curve
steeper than that of the SMC if they have blue intrinsic UV slopes.
In addition, as noted in Section~\ref{sec:irxmass}, galaxies in our
low-mass ($M_\ast \le 10^{9.75}$\,$M_\odot$) subsample appear to also
require a dust curve steeper than that of the SMC.

Figure~\ref{fig:irxyoung} summarizes a few recent measurements of
IRX-$\beta$ for young and low-mass galaxies at $z\sim 2$
\citep{baker01, siana09, reddy10, reddy12a}, low-mass galaxies and
LBGs in general at $z\sim 4-10$ \citep{bouwens16b}, and our own
measurements.  The compilation from \citet{reddy10} and
\citet{reddy12a} includes 24\,$\mu$m constraints on the IRX of young
galaxies.  Shifting their IRX by $\approx 0.35$\,dex to higher values
to account for the deficit of PAH emission in galaxies with $\la
400$\,Myr \citep{shivaei16} results in upper limits or a stacked
measurement of IRX that are broadly consistent with either the SMC
curve or one that is steeper.

Considering the {\em Herschel} measurements here and in
\citet{reddy12a}, and ALMA measurements at $z\sim 2-10$
\citep{capak15, bouwens16b}, we find that young/low-mass galaxies at
$z\ga 2$ follow a dust curve steeper than that of the SMC,
particularly in the context of a blue intrinsic slope, $\beta_0 =
-2.62$.  Note that unlike cB58, which has a stellar metallicity of
$\simeq 0.25$\,$Z_\odot$ \citep{pettini00}, the Cosmic Eye has a
metallicity of $\sim 0.5$\,$Z_\odot$, suggesting a relatively red
intrinsic UV slope.  In this case, the IRX of the Cosmic Eye may be
described adequately by the SMC curve.  There is additional evidence
of suppressed IRX ratios at lower mass from rest-optical emission line
studies of $z\sim 2$ galaxies.  In particular, \citet{reddy15} found
that a significant fraction of $z\sim 2$ galaxies with $M_{\ast}\la
10^{9.75}$\,$M_\odot$ have very red $\beta$, or $\ebmv$, relative to
the reddening deduced from the Balmer lines (e.g., see their
Figure~17), implying that such galaxies would have lower IRX for a
given $\beta$ than that predicted by common attenuation/extinction
curves.

Evidence for curves steeper than the SMC average have been observed
along certain sightlines within the SMC \citep{gordon03}, in the Milky
Way and some Local Group galaxies \citep{gordon03, sofia05, gordon09,
  amanullah14}, and in quasars (e.g., \citealt{hall02, jiang13,
  zafar15}).  Our results, combined with those in the literature,
suggest that such steep curves may be typical of low-mass and young
galaxies at high redshift.  While the attenuation curve will
undoubtedly vary from galaxy-to-galaxy depending on the star-formation
history, age, metallicity, dust composition, and geometrical
configuration of stars and dust, the fact that young/low-mass galaxies
lie {\em systematically} below the IRX-$\beta$ relation predicted with
an SMC curve suggests that a steep curve may apply uniformly to such
galaxies.  

An unresolved issue is the physical reason why young and low-mass
galaxies may follow a steeper attenuation curve than their older and
more massive counterparts.  \citet{reddy12a} suggest a possible
scenario in which galaxies transition from steep to gray attenuation
curves as they age due to star formation occurring over extended
regions and/or the cumulative effects of galactic outflows that are
able to clear the gas and dust along certain sightlines.
On the other hand, if young and low-mass galaxies have higher ionizing
escape fractions as a result of lower gas/dust covering fractions
(e.g., \citealt{reddy16b}),
then one might expect their attenuation curve to exhibit a shallower
dependence on wavelength than the SMC extinction curve.  

In any case,
curves steeper than the SMC may be possible with a paucity of large
dust grains and/or an over-abundance of silicate grains
\citep{zafar15}.  In particular, large dust grains may be efficiently
destroyed by SN shock waves \citep{draine79, mckee89, jones04}, which
would have the effect of steepening the dust curve (i.e., such that
proportionally more light is absorbed in the UV relative to the
optical).
If the destruction of large grains is significant in young/low-mass
galaxies, then it may explain both their red $\beta$ and their low
IRX.  Alternatively, the lower gas-phase [Si/O] measured from the
composite rest-frame UV spectrum of $z\sim 2$ galaxies relative to the
solar value indicates significant depletion of Si onto dust grains,
while carbon is under-abundant relative to oxygen \citep{steidel16}.
This result may suggest an enhancement of silicate over carbonaceous
grains that may result in a steeper attenuation curve.

\begin{figure*}
\epsscale{1.0}
\plotone{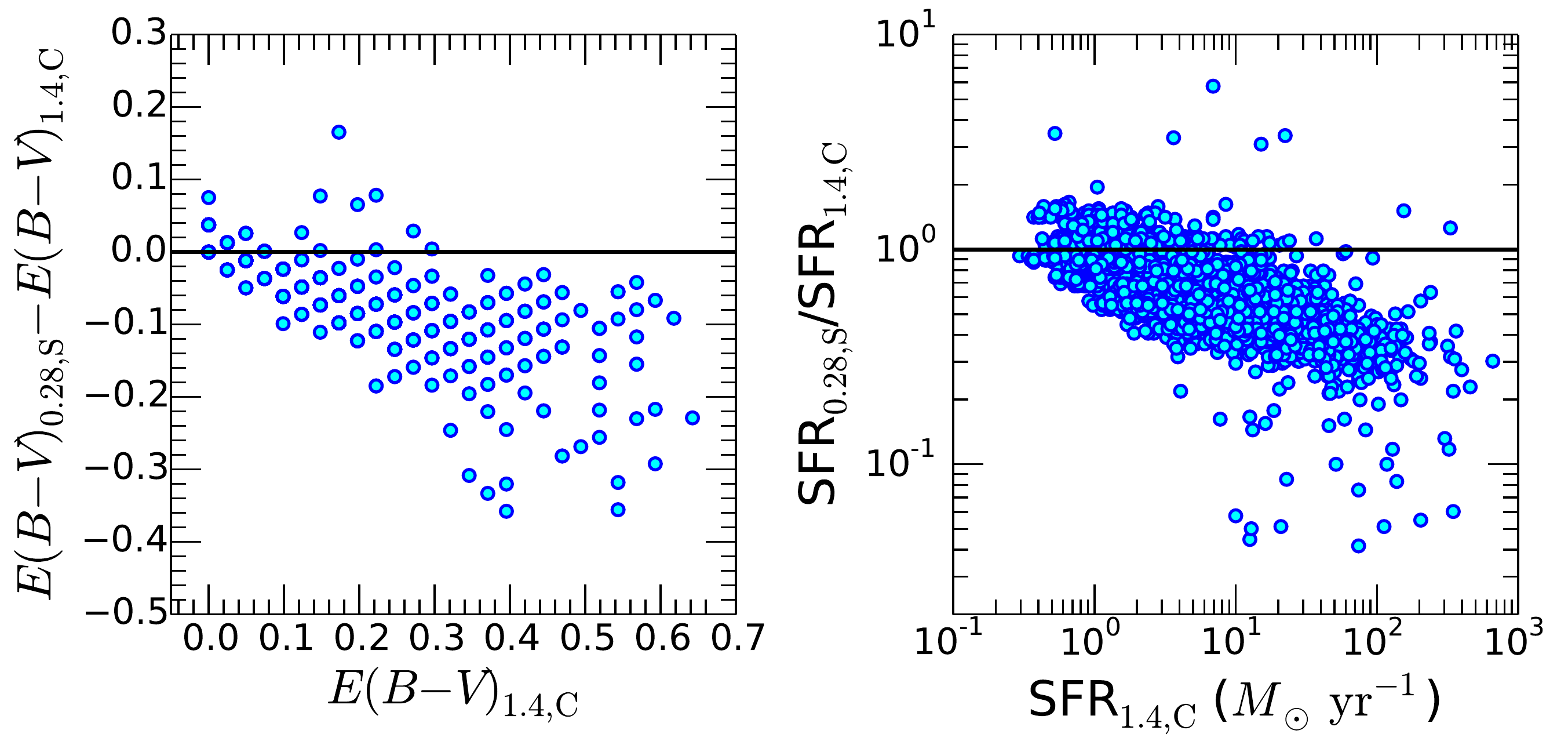}
\caption{ Comparison of the derived reddening, $\ebmv$, and SFRs when
  considering the cases where (a) $1.4Z_\odot$ BC03 model with the
  \citet{calzetti00} attenuation curve (``1.4,C'' subscripts) and (b)
  $0.28Z_\odot$ BC03 model with SMC extinction curve (``0.28,S'')
  subscripts.  The typical random uncertainties are $\delta \ebmv
  \simeq 0.04$ and $\delta \log[{\rm SFR}/M_\odot\,{\rm yr}^{-1}]
  \simeq 0.17$\,dex.  The sparse sampling in the left-hand panel is
  due to the fixed increments of $\delta A_{\rm V}=0.1$ used when
  fitting the SEDs of galaxies in our sample.  The difference in
  reddening between the two cases becomes increasingly discrepant for
  objects with redder colors.  A lower $\ebmv$ is required to
  reproduce an observed UV slope given the steep far-UV rise of the
  SMC extinction curve compared to that of the \citet{calzetti00}
  curve.  These differences in reddening, combined with the particular
  values of the SMC and \citet{calzetti00} attenuation curves at
  $\lambda = 1600$\,\AA, result in SFRs that are generally lower in
  case (b) than in case (a).}
\label{fig:sedcompare}
\end{figure*}

\subsection{Implications for Stellar Populations and Ionizing Production Efficiencies}

Inferring the intrinsic stellar populations of galaxies based on their
observed photometry requires one to adopt some form of the dust
attenuation curve.  It is therefore natural to ask whether these
inferences change in light of our findings of a steeper (SMC-like)
attenuation curve for $z\sim 2$ galaxies with intrinsically blue UV
spectral slopes.  To address this issue, we re-modeled (using FAST)
the SEDs of galaxies in our sample assuming two cases: (a) a
$1.4Z_\odot$ BC03 model (canonically referred to as the ``solar''
metallicity model using old solar abundance measurements) with the
\citet{calzetti00} attenuation curve; and (b) a $0.28Z_\odot$ BC03
model with an SMC extinction curve.  The ages derived in case (b) are
on average $30\%$ older than those derived in case (a), primarily
because less reddening is required to reproduce a given UV slope with
the SMC extinction curve, resulting in older ages (e.g., see
discussion in \citealt{reddy12b}).  Similarly, the stellar masses
derived in case (b) are on average $30\%$ lower than those derived in
case (a).

Perhaps most relevant in the context of our analysis are the changes
in inferred reddening and SFR.  As shown in
Figure~\ref{fig:sedcompare}, the reddening deduced from the SMC
extinction curve is lower than that obtained with \citet{calzetti00}
attenuation curve, owing to the steep rise in the far-UV of the former
relative to the latter.  The largest differences in $\ebmv$ and SFR
occur for the reddest objects and those with larger SFRs.  We find
the following relations between the reddening and SFRs derived for the
two cases discussed above:
\begin{eqnarray}
\ebmv_{0.28Z_\odot,{\rm SMC}} = 0.65\ebmv_{1.4Z_\odot,{\rm Calz}}
\end{eqnarray}
and
\begin{eqnarray}
\log[{\rm SFR}_{0.28Z_\odot,{\rm SMC}}/M_\odot\,{\rm yr}^{-1}] = \nonumber \\
0.79 \log[{\rm SFR}_{1.4Z_\odot,{\rm Calz}}/M_\odot\,{\rm yr}^{-1}] -0.05.
\end{eqnarray}
The lower SFRs derived in the SMC case result in a factor of $\approx
2$ lower SFR densities at $z\ga 2$, as discussed in some depth in
\citet{bouwens16b}.  

The general applicability of an SMC-like dust curve to high-redshift
galaxies is also of particular interest when considering its impact on
the ionizing efficiencies of such galaxies, a key input to
reionization models \citep{robertson13, bouwens15b, bouwens16a}.  Specifically,
the ionizing photon production efficiency, $\xi_{\rm ion}$, is simply
the ratio of the rate of H-ionizing photons produced to the
non-ionizing UV continuum luminosity.  This quantity is directly
related to another commonly-used ratio, namely the Lyman continuum
flux density (e.g., at $800$ or $900$\,\AA) to the non-ionizing UV
flux density (e.g., at $1600$\,\AA), $f_{\rm 900}/f_{\rm 1600}$.  

The ionizing photon production efficiency is typically computed by
combining Balmer emission lines, such as H$\alpha$, with UV continuum
measurements (e.g., \citealt{bouwens15b}), but both the lines and the
continuum must be corrected for dust attenuation.  In the context of
our present analysis, the dust corrections for the UV continuum are
lower by factors of $1.12$, $1.82$, and $2.95$, for galaxies with
Calzetti-inferred SFRs of 1, 10, and 100\,$M_\odot$\,yr$^{-1}$,
respectively.  Thus, for given Balmer line luminosities that are
corrected for dust using the Galactic extinction curve
\citep{calzetti94, steidel14, reddy15}, $\xi_{\rm ion}$ would be
correspondingly larger by the same factors by which the dust-corrected
UV luminosities are lower.

A secondary effect that will boost $\xi_{\rm ion}$ above the
predictions from single star solar metallicity stellar population
models is the higher ionizing photon output associated with lower
metallicity and rotating massive stars \citep{eldridge09,
  leitherer14}.  In particular, the fiducial $0.14Z_\odot$ BPASS model
that includes binary evolution and an IMF extending to 300\,$M_\odot$
predicts a factor of $\approx 3$ larger H$\alpha$ luminosity per solar
mass of star formation after 100\,Myr of constant star formation
relative to that computed using the \citet{kennicutt12} relation,
which assumes a solar metallicity Starburst99 model.  On the other
hand, the UV luminosity is larger by $\approx 30\%$.  Thus, such
models predict a $\xi_{\rm ion}$ that is elevated by a factor of
$\approx 2$ relative to those assumed in standard calibrations between
H$\alpha$/UV luminosity and SFR (e.g., see also \citealt{nakajima16,
  bouwens16a, stark15, reddy16b}).  Consequently, calculations or
predictions of $\xi_{\rm ion}$ for high-redshift galaxies should take
into account the effects of a steeper attenuation curve and lower
metallicity stellar populations that may include stellar
rotation/binarity.  {\em Our results suggest that an elevated value of
  $\xi_{\rm ion}$ is not only a feature of very high-redshift ($z\ga
  6$) galaxies, but may be quite typical for $z\sim 2$ galaxies as
  well.}


\section{CONCLUSIONS}

In this paper, we have presented an analysis of the relationship
between dust obscuration (IRX$=L_{\rm IR}/L_{\rm UV}$) and other
commonly-derived galaxy properties, including UV slope ($\beta$),
stellar mass ($M_{\ast}$), and UV luminosity ($L_{\rm UV}$), for a
large sample of 3,545 rest-optically selected star-forming galaxies at
$z=1.5-2.5$ drawn from HDUV UVIS and 3D-HST photometric catalogs of
the GOODS fields.  Our sample is unique in that it significantly
extends the dynamic range in $\beta$ and $L_{\rm UV}$ compared to
previous UV-selected samples at these redshifts.  In particular, close
to $60\%$ of the objects in our sample have UV slopes bluer than
$\beta = -1.70$ and $>95\%$ have rest-frame UV absolute magnitudes
fainter than the characteristic magnitude at these redshifts, with the
faintest galaxies having $L_{\rm UV}\approx 0.05L^{\ast}_{\rm UV}$.

We use stacks of the deep {\em Herschel}/PACS imaging in the GOODS
fields to measure the average dust obscuration for galaxies in our
sample and compare it to predictions of the IRX-$\beta$ relation for
different stellar population models and attenuation/extinction curves
using energy balance arguments.  Specifically, we consider the
commonly adopted \citet{bruzual03} stellar population models for
different metallicities ($0.28$ and $1.4Z_\odot$), as well as the
low-metallicity ($0.14Z_\odot$) BPASS model.  Additionally, we compute
predictions of the IRX-$\beta$ relations for the \citet{calzetti00}
and \citet{reddy15} dust attenuation curves, and the SMC extinction
curve.  The lower metallicity stellar population models result in
significant shifts in the IRX-$\beta$ relation of up to $\delta\beta =
0.4$ towards bluer $\beta$ relative to the canonical relation of
\citet{meurer99}.  In the context of the lower metallicity stellar
population models applicable for high-redshift galaxies, we find that
the strong trend between IRX and $\beta$ measured from the HDUV and
3D-HST samples follows most closely that predicted by the SMC
extinction curve.  We find that grayer attenuation curves (e.g.,
\citealt{calzetti00}) over-predict the IRX at a given $\beta$ by a
factor of $\ga 3$ when assuming intrinsically blue UV spectral slopes.
{\em Thus, our results suggest that an SMC curve is the one most
applicable to lower stellar metallicity populations at high redshift.}

Performing a complementary stacking analysis of the {\em Spitzer}/MIPS
$24$\,$\mu$m images implies an average mid-IR-to-IR luminosity ratio,
$\langle L_{7.7}/L_{\rm IR}\rangle$, that is a factor of $3-4$ lower
than for galaxies with reddest ($\beta>-0.5$) and the UV-brightest
($M_{1600}\la -21$) and UV-faintest ($M_{1600}\ga -19$) galaxies
relative to the average for all galaxies in our sample
(Appendix~\ref{sec:l8lir}).  These results indicate large variations
in the conversion between rest-frame 7.7\,$\mu$m and IR luminosity.

At any given UV luminosity, galaxies with redder $\beta$ have larger
IRX.  IRX-$L_{\rm UV}$ relations for blue and red star-forming
galaxies average together to result in a roughly constant IRX of
$\simeq 3-4$ over roughly two decades in UV luminosity ($2\times 10^9
\la L_{\rm UV}\la 2\times 10^{11}$\,$L_\odot$).  Consequently, the
bluer $\beta$ observed for UV-faint galaxies seen in this work and
previous studies may simply reflect intrinsically bluer UV spectral
slopes for such galaxies, rather than signifying changes in the dust
obscuration.

Galaxies with stellar masses $M_{\ast}>10^{9.75}$\,$M_\odot$ have an
IRX-$\beta$ relation that is consistent with the SMC extinction curve,
while the lower mass galaxies in our sample with $M_{\ast}\le
10^{9.75}$\,$M_\odot$ have an IRX-$\beta$ relation that is {\em at
  least} as steep as the SMC.  The shifting of the IRX-$\beta$
relations towards bluer $\beta$ for the lower metallicity stellar
populations expected for high-redshift galaxies implies that the
low-mass galaxies in our sample, as well as the low-mass and young
galaxies from previous studies, require a dust curve steeper than that
of the SMC.  The low metallicity stellar populations favored for
high-redshift galaxies result in steeper attenuation curves and higher
ionizing photon production rates which, in turn, facilitate the role
that galaxies may have in reionizing the universe at very high
redshift or keeping the universe ionized at lower redshifts ($z\sim
2$).

There are several future avenues for building upon this work.
Detailed spectral modeling of the rest-UV and/or rest-optical spectra
of galaxies (e.g., \citealt{steidel16, reddy16b}) may be used to
discern their intrinsic spectral slopes and thus disentangle the
effects of the intrinsic $\beta$ and the attenuation curve relevant
for high-redshift galaxies.  Second, the higher spatial resolution and
depth of X-ray observations (compared to the far-IR) makes them
advantageous for investigating the bolometric SFRs and hence dust
obscuration for galaxies substantially fainter than those directly
detected with either {\em Spitzer} or {\em Herschel} in reasonable
amounts of time, provided that the scaling between X-ray luminosity
and SFR can be properly calibrated (e.g., for metallicity effects;
\citealt{basuzych13}).  In addition, nebular recombination line
estimates of dust attenuation (e.g., from the Balmer decrement;
\citealt{reddy15}) may be used to assess the relationship between IRX
and $\beta$ for individual star-forming galaxies, rather than through
the stacks necessitated by the limited sensitivity of far-IR imaging.

\acknowledgements

This work was supported by NASA through grant HST-GO-13871 and from
the Space Telescope Science Institute, which is operated by AURA,
Inc., under NASA contract NAS 5-26555.  K. Penner kindly provided data
from his published work in electronic format.  NAR is supported by an
Alfred P. Sloan Research Fellowship.




\appendix

\section{A. $\ebmv$ vs. $\beta$ and IRX vs. $\beta$ Relations}
\label{sec:sumrelations}

In Table~\ref{tab:sumrelations}, we summarize the relations between
$\beta$ and $\ebmv$ and between IRX and $\beta$ for different
assumptions of the dust curve, heating from Ly$\alpha$, inclusion of
nebular continuum emission, and the normalization of the dust curve.

\begin{deluxetable}{llcccccc}[!h]
\tabletypesize{\footnotesize}
\tablewidth{0pc}
\tablecaption{Summary of $\beta$ vs. $\ebmv$ and IRX vs. $\beta$ Relations}
\tablehead{
\colhead{} & 
\colhead{} & 
\colhead{Nebular} &
\colhead{Ly$\alpha$} & 
\colhead{} &
\colhead{} &
\colhead{} &
\colhead{} \\
\colhead{Model\tablenotemark{a}} &
\colhead{Dust Curve\tablenotemark{b}} &
\colhead{Continuum\tablenotemark{c}} &
\colhead{Heating\tablenotemark{d}} &
\colhead{$\delta R_{\rm V}$\tablenotemark{e}} &
\colhead{$\beta$\tablenotemark{f}} &
\colhead{} &
\colhead{IRX}}
\startdata
BPASS - $0.14Z_\odot$ & Calzetti+00 & Yes & Yes & 0.0 & $-2.616 + 4.684\ebmv$ & & $1.67\times[10^{0.4(2.13\beta+5.57)}-1]$ \\
                     &             & Yes & Yes & 1.5 &                       & & $1.66\times[10^{0.4(1.81\beta+4.73)}-1]$ \\
                     &             & Yes & No & 0.0 &                        & & $1.39\times[10^{0.4(2.13\beta+5.57)}-1]$ \\
                     &             & Yes & No & 1.5 &                        & & $1.38\times[10^{0.4(1.81\beta+4.73)}-1]$ \\
                     &             & No & Yes & 0.0 & $-2.709 + 4.684\ebmv$ & & $1.73\times[10^{0.4(2.13\beta+5.76)}-1]$ \\
                     &             & No & Yes & 1.5 &                       & & $1.73\times[10^{0.4(1.81\beta+4.90)}-1]$ \\
                     &             & No & No & 0.0 &                        & & $1.44\times[10^{0.4(2.13\beta+5.76)}-1]$ \\
                     &             & No & No & 1.5 &                        & & $1.42\times[10^{0.4(1.81\beta+4.90)}-1]$ \\
\\
                     & Reddy+15    & Yes & Yes & 0.0 & $-2.616 + 4.594\ebmv$ & & $1.68\times[10^{0.4(1.82\beta+4.77)}-1]$ \\
                     &             & Yes & Yes & 1.5 &                       & & $1.67\times[10^{0.4(1.50\beta+3.92)}-1]$ \\
                     &             & Yes & No & 0.0 &                        & & $1.40\times[10^{0.4(1.82\beta+4.77)}-1]$ \\
                     &             & Yes & No & 1.5 &                        & & $1.38\times[10^{0.4(1.50\beta+3.92)}-1]$ \\
                     &             & No & Yes & 0.0 & $-2.709 + 4.594\ebmv$ & & $1.74\times[10^{0.4(1.82\beta+4.94)}-1]$ \\
                     &             & No & Yes & 1.5 &                       & & $1.74\times[10^{0.4(1.50\beta+4.05)}-1]$ \\
                     &             & No & No & 0.0 &                        & & $1.44\times[10^{0.4(1.82\beta+4.94)}-1]$ \\
                     &             & No & No & 1.5 &                        & & $1.43\times[10^{0.4(1.50\beta+4.05)}-1]$ \\
\\
                     & SMC (Gordon+03) & Yes & Yes & 0.0 & $-2.616 + 11.259\ebmv$ & & $1.79\times[10^{0.4(1.07\beta+2.79)}-1]$ \\
                     &             & Yes & Yes & 1.5 &                           & & $1.80\times[10^{0.4(0.93\beta+2.44)}-1]$ \\
                     &             & Yes & No & 0.0 &                            & & $1.47\times[10^{0.4(1.07\beta+2.79)}-1]$ \\
                     &             & Yes & No & 1.5 &                            & & $1.47\times[10^{0.4(0.93\beta+2.44)}-1]$ \\
                     &             & No & Yes & 0.0 & $-2.709 + 11.259\ebmv$ & & $1.83\times[10^{0.4(1.07\beta+2.89)}-1]$ \\
                     &             & No & Yes & 1.5 &                        & & $1.85\times[10^{0.4(0.93\beta+2.52)}-1]$ \\
                     &             & No & No & 0.0 &                         & & $1.50\times[10^{0.4(1.07\beta+2.89)}-1]$ \\
                     &             & No & No & 1.5 &                        & & $1.50\times[10^{0.4(0.93\beta+2.52)}-1]$ \\
\\
BC03 - $1.4Z_\odot$ & Calzetti+00 & Yes & Yes & 0.0 & $-2.383 + 4.661\ebmv$ & & $1.44\times[10^{0.4(2.14\beta+5.10)}-1]$ \\
                     &             & Yes & Yes & 1.5 &                       & & $1.41\times[10^{0.4(1.82\beta+4.33)}-1]$ \\
                     &             & Yes & No & 0.0 &                        & & $1.35\times[10^{0.4(2.14\beta+5.10)}-1]$ \\
                     &             & Yes & No & 1.5 &                        & & $1.32\times[10^{0.4(1.82\beta+4.33)}-1]$ \\
                     &             & No & Yes & 0.0 & $-2.439 + 4.661\ebmv$ & & $1.47\times[10^{0.4(2.14\beta+5.22)}-1]$ \\
                     &             & No & Yes & 1.5 &                       & & $1.44\times[10^{0.4(1.82\beta+4.43)}-1]$ \\
                     &             & No & No & 0.0 &                        & & $1.38\times[10^{0.4(2.14\beta+5.22)}-1]$ \\
                     &             & No & No & 1.5 &                        & & $1.35\times[10^{0.4(1.82\beta+4.43)}-1]$ \\
\\
                     & Reddy+15    & Yes & Yes & 0.0 & $-2.383 + 4.568\ebmv$ & & $1.43\times[10^{0.4(1.83\beta+4.37)}-1]$ \\
                     &             & Yes & Yes & 1.5 &                       & & $1.39\times[10^{0.4(1.51\beta+3.59)}-1]$ \\
                     &             & Yes & No & 0.0 &                        & & $1.40\times[10^{0.4(1.82\beta+4.77)}-1]$ \\
                     &             & Yes & No & 1.5 &                        & & $1.30\times[10^{0.4(1.51\beta+3.59)}-1]$ \\
                     &             & No & Yes & 0.0 & $-2.439 + 4.568\ebmv$ & & $1.46\times[10^{0.4(1.83\beta+4.47)}-1]$ \\
                     &             & No & Yes & 1.5 &                       & & $1.42\times[10^{0.4(1.51\beta+3.67)}-1]$ \\
                     &             & No & No & 0.0 &                        & & $1.37\times[10^{0.4(1.83\beta+4.47)}-1]$ \\
                     &             & No & No & 1.5 &                        & & $1.33\times[10^{0.4(1.51\beta+3.67)}-1]$ \\
\\
                     & SMC (Gordon+03) & Yes & Yes & 0.0 & $-2.383 + 11.192\ebmv$ & & $1.47\times[10^{0.4(1.07\beta+2.55)}-1]$ \\
                     &             & Yes & Yes & 1.5 &                           & & $1.45\times[10^{0.4(0.94\beta+2.23)}-1]$ \\
                     &             & Yes & No & 0.0 &                            & & $1.38\times[10^{0.4(1.07\beta+2.55)}-1]$ \\
                     &             & Yes & No & 1.5 &                            & & $1.35\times[10^{0.4(0.94\beta+2.23)}-1]$ \\
                     &             & No & Yes & 0.0 & $-2.439 + 11.192\ebmv$ & & $1.50\times[10^{0.4(1.07\beta+2.61)}-1]$ \\
                     &             & No & Yes & 1.5 &                        & & $1.49\times[10^{0.4(0.94\beta+2.29)}-1]$ \\
                     &             & No & No & 0.0 &                         & & $1.40\times[10^{0.4(1.07\beta+2.61)}-1]$ \\
                     &             & No & No & 1.5 &                        & & $1.38\times[10^{0.4(0.94\beta+2.29)}-1]$ \\
\\
Meurer+99            & ---         & --- & --- & --- & $-2.23 + 5.01\ebmv$ & & $2.07\times[10^{0.4(1.99\beta+4.43)}-1]$
\enddata
\tablenotetext{a}{Stellar population model, assuming a constant star-formation history
and an age of 100\,Myr.  For reference, we also list the original relations of \citet{meurer99}, where their IRX-$\beta$ relation is shifted to higher IRX by $0.24$\,dex (see text).}
\tablenotetext{b}{Dust attenuation curves from \citet{calzetti00} and \citet{reddy15} and
the dust extinction curve for the SMC from \citet{gordon03}, where all have been updated
for the shape of the curve at $\lambda \la 1200$\,\AA\, according to the procedure of
\citet{reddy16a}.}
\tablenotetext{c}{Indicates whether the nebular continuum emission is included.}
\tablenotetext{d}{Indicates whether dust heating by Ly$\alpha$ photons is included.}
\tablenotetext{e}{Indicates the value by which the dust attenuation curves are shifted
{\em downward} in normalization ($R_{\rm V}$).}
\tablenotetext{f}{A blank entry indicates that the relation between $\beta$ and $\ebmv$
is the same as that of the previous non-blank entry in this column.}

\label{tab:sumrelations}
\end{deluxetable}

\section{B. Inferences of IRX from {\em Spitzer}/MIPS $24$\,$\mu$m Data}
\label{sec:l8lir}

Prior to the launch of {\em Herschel}, most non-UV-based inferences of
the dust content of $L^{\ast}$ galaxies at $z\ga 1.5$ relied on the
detection of the redshifted mid-IR emission bands, commonly associated
with PAHs, with the {\em Spitzer} MIPS instrument.  A number of early
studies of local and $z\sim 1$ galaxies suggested that PAH emission
correlates with total dust emission (e.g., \citealt{roussel01,
  forster03, forster04}), though with some variations with
metallicity, ionizing intensity, star-formation-rate surface density,
and stellar population age (e.g., \citealt{normand95, helou01,
  alonso04, engelbracht05, hogg05, madden06, draine07b, smith07,
  galliano08, hunt10, sales10, elbaz11, magdis13, seok14}).  Recently,
\citet{shivaei16} presented the first statistically significant trends
showing lower ratios of the $7.7$\,$\mu$m-to-total IR luminosity,
$L_{\rm 7.7}/L_{\rm IR}$, with higher ionization intensity, lower
gas-phase metallicity, and younger ages for galaxies at $z\sim 2$ from
the MOSFIRE Deep Evolution Field Survey \citep{kriek15}.  The
aforementioned studies have suggested either a delayed enrichment of
PAHs in young galaxies or the destruction of PAHs in high ionization
and low metallicity environments.

\begin{figure*}
\epsscale{1.0}
\plotone{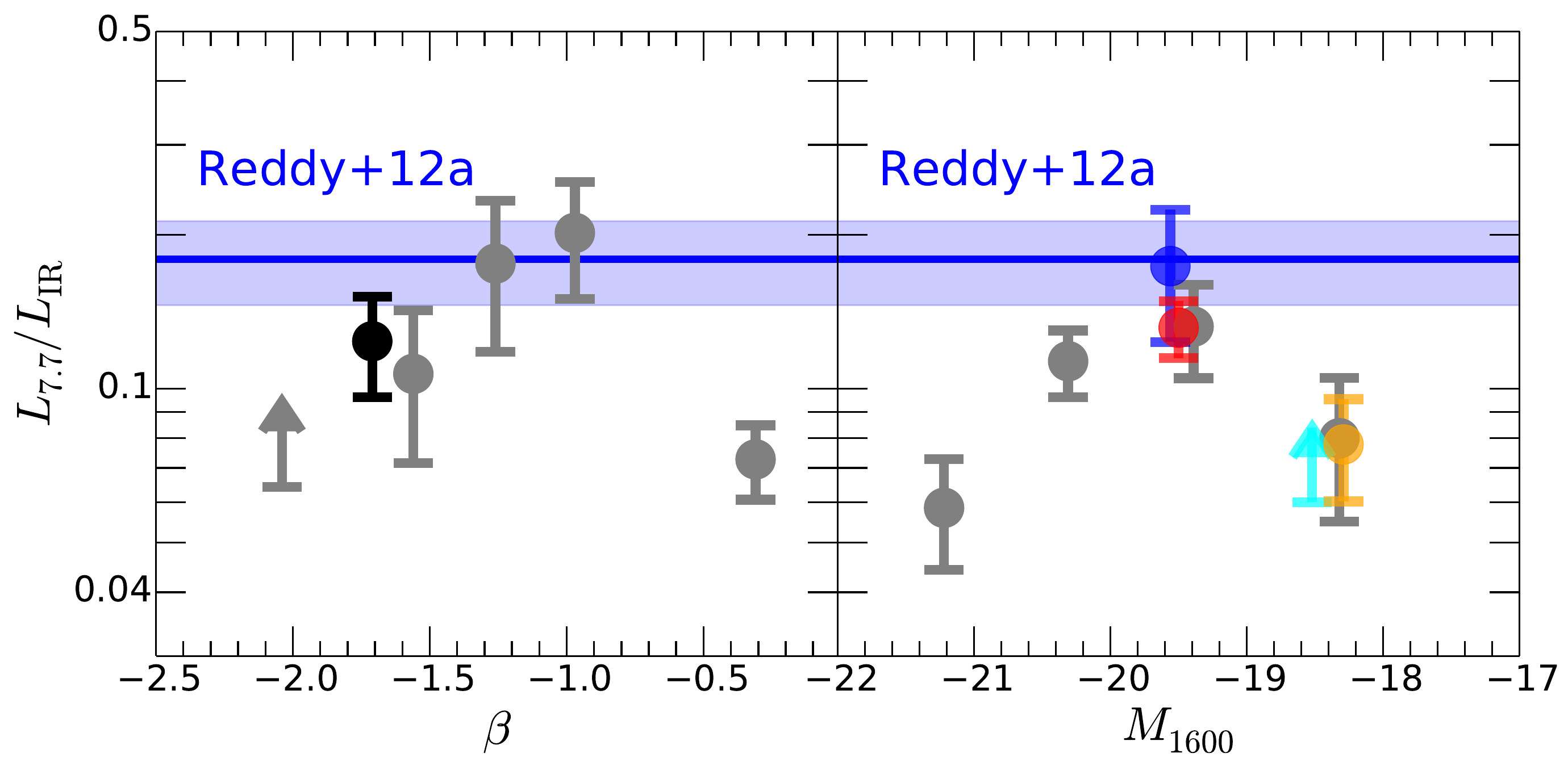}
\caption{Ratio of $7.7$\,$\mu$m-to-total IR luminosity as a function
  of UV slope and absolute magnitude at $1600$\,\AA.  The gray points
  indicate the ratios measured in bins of $\beta$ (left) and
  $M_{1600}$ (right).  The ratio measured for the entire sample is
  indicated by the black point in the left panel.  In the right panel,
  the blue and red points indicate the results for UV-bright galaxies
  with $M_{1600}\le -19$ and $\beta \le -1.4$ and $\beta>-1.4$,
  respectively.  Similarly, the cyan and orange points indicate the
  values for UV-faint galaxies with $M_{1600}>-19$ and $\beta \le
  -1.4$ (offset slightly for clarity) and $\beta>-1.4$, respectively.
  The mean and $1\sigma$ uncertainty in $L_{7.7}/L_{\rm IR}$ for a
  sample of 124 UV-selected galaxies at $z\sim 2$ from
  \citet{reddy12a} are indicated by the blue line and shaded region,
  respectively.}
\label{fig:l8lir}
\end{figure*}

In light of these findings, we considered $L_{\rm 7.7}/L_{\rm IR}$ for
the predominantly blue, low luminosity galaxies in our
sample.\footnote{As virtually all of the galaxies in our sample are
  undetected in the PACS imaging, we were not able to normalize the
  $24$\,$\mu$m images by $1/L_{\rm IR}$ before stacking them (e.g., in
  the same way that we were able to normalize them by $1/L_{\rm UV}$;
  Section~\ref{sec:stacking}).  However, as the stacked fluxes are
  weighted by $1/L_{\rm UV}$, and $L_{\rm UV}\propto SFR\propto L_{\rm
    IR}$ for all but the dustiest galaxies in our sample
  (Section~\ref{sec:irxuvlum}), we assumed that the ratio of the
  average luminosities is similar to the average ratio of the
  luminosities, i.e., $\langle L_{\rm 7.7}\rangle/\langle L_{\rm
    IR}\rangle \approx \langle L_{\rm 7.7}/L_{\rm IR}\rangle$ (see
  discussion in \citealt{shivaei16}).  Thus, we simply divided $L_{\rm
    7.7}$ by $L_{\rm IR}$ for each stack presented in
  Table~\ref{tab:stackedresults} in order to deduce the average ratio
  $\langle L_{\rm 7.7}/L_{\rm IR}\rangle$.}  The galaxies in our
sample as a whole exhibit $\langle L_{\rm 7.7}/L_{\rm IR}\rangle =
0.12\pm0.03$, similar within $1\sigma$ to that computed in
\citet{reddy12a}, $\langle L_{\rm 7.7}/L_{\rm IR}\rangle =
0.18\pm0.03$, where the latter has been corrected for the difference
in this ratio when assuming the \citet{elbaz11} dust template rather
than those of \citet{chary01}.  However, when divided into bins of UV
slope, we find that $\langle L_{\rm 7.7}/L_{\rm IR}\rangle$ is
substantially lower for galaxies with the reddest $\beta$, as well as
for the brightest ($M_{1600}\le -21$) and faintest ($M_{1600} > -19$)
galaxies in our sample (Figure~\ref{fig:l8lir}).  

The low ratio ($\langle L_{\rm 7.7}/L_{\rm IR}\rangle = 0.07\pm0.01$)
observed for galaxies with the reddest UV slopes may be related to
significant $9.9$\,$\mu$m silicate absorption affecting the observed
$24$\,$\mu$m flux.  An alternative explanation invokes an IR
luminosity that is boosted in the presence of AGN, though we consider
this possibility unlikely as we have removed obscured AGN from the
sample based on their IRAC colors (Section~\ref{sec:sample}).
Regardless, the lower $\langle L_{\rm 7.7}/L_{\rm IR}\rangle$ found
for these red galaxies is partly responsible for a similar low ratio
found for the faintest galaxies in our sample, since many of the
dustiest galaxies in our sample are also UV-faint
(Figure~\ref{fig:zmag}).  Isolating the UV-faint galaxies with slopes
bluer than $\beta=-1.4$ yields an unconstraining lower limit on
$\langle L_{\rm 7.7}/L_{\rm IR}\rangle$.  

Finally, we note that the brightest galaxies in our sample
($M_{1600}\le -21$) also exhibit a very low $\langle L_{\rm
  7.7}/L_{\rm IR}\rangle = 0.06\pm0.01$.  Such galaxies are on average
a factor of $1.6\times$ younger than $M_{1600}>-21$ galaxies ($\approx
500$ vs. $\approx 800$\,Myr).  Thus, their lower mean ratio may be
related to a deficit of PAHs due to younger stellar population ages,
or may be related to harder ionization fields and/or lower gas-phase
metallicities.  Unfortunately, we are unable to fully explore how the
$L_{7.7}$-to-$L_{\rm IR}$ ratio varies with age given unconstraining
lower limits on this value for galaxies with ages $\la 500$\,Myr.
Irrespective of the physical causes for changes in the PAH-to-infrared
luminosity ratio, our results suggest that caution must be used when
adopting a single-valued conversion to recover $L_{\rm IR}$ from
mid-IR measurements.  For example, assuming the mean ratio found for
our sample would result in $24$\,$\mu$m-inferred $L_{\rm IR}$ that are
a factor of $\approx 2$ lower than the ``true'' values for the reddest
($\beta >-0.8$) and UV-brightest ($M_{1600}\le -21$) galaxies in our
sample.





\end{document}